\documentclass[aps,reprint,prb,letterpaper,superscriptaddress]{revtex4-2}

\pdfoutput=1

\usepackage{amsmath,amssymb,amsthm,mathrsfs, cancel}
\usepackage{siunitx,mhchem,nicefrac}

\usepackage{miller}
\usepackage{physics}

\usepackage{graphicx}
\usepackage{wrapfig, subfig}

\usepackage{tabularx, booktabs}

\usepackage{blindtext}
\usepackage{array,dcolumn}

\usepackage[all]{nowidow}
\usepackage{hyperref}
\hypersetup{
	colorlinks=true,
	linkcolor=blue,
	citecolor=blue,
	urlcolor=blue,
}

\usepackage{lipsum}
\usepackage{xcolor}

\begin{document}
	
\title{Incorporation of random alloy \ce{GaBi_xAs_{1-x}} barriers in InAs quantum dot molecules: alloy strain and orbital effects towards enhanced tunneling}
	
\author{Arthur Lin}
\email{arthur.lin@nist.gov}
\affiliation{Joint Quantum Institute, \\
	University of Maryland and National Institute of Standards and Technology,
	College Park, Maryland 20742, USA
}

\author{Matthew F.\ Doty}
\affiliation{Department of Materials Science and Engineering, 
	University of Delaware,
	Newark, Deleware 19716, USA
}
	
\author{Garnett W.\ Bryant}
\affiliation{Joint Quantum Institute, \\
	University of Maryland and National Institute of Standards and Technology,
	College Park, Maryland 20742, USA
}
\affiliation{Nanoscale Device Characterization Division, \\
	National Institute of Standards and Technology,
	Gaithersburg, Maryland 20899, USA
}

\date{\today}
	
	
\begin{abstract}
	Self-assembled InAs quantum dots (QDs), which have long hole-spin coherence times and are amenable to optical control schemes, have long been explored as building blocks for qubit architectures.  One such design consists of vertically stacking two QDs to create a quantum dot molecule (QDM).  The two dots can be resonantly tuned to form "molecule-like" coupled hole states from the hybridization of hole states otherwise localized in each respective dot.  Furthermore, spin-mixing of the hybridized states in dots offset along their stacking direction enables qubit rotation to be driven optically, allowing for an all-optical qubit control scheme. Increasing the magnitude of this spin mixing is important for optical quantum control protocols. To enhance the tunnel coupling and spin-mixing across the dots, we introduce Bi in the \ce{GaAs} inter-dot barrier.  Previously, we showed how to model \ce{InAs}/\ce{GaBi_xAs_{1-x}} in an atomistic tight-binding formalism, and how the dot energy levels are affected by the alloy.  In this paper, we discuss the lowering of the tunnel barrier, which results in a three fold increase of hole tunnel coupling strength in the presence of a 7\% alloy.  Additionally, we show how an asymmetric strain between the two dots caused by the alloy shifts the resonance.  Finally, we discuss device geometries for which the introduction of Bi is most advantageous.
\end{abstract}
	
\maketitle
	
%%%%%%%%%%%%%%%%%%%%%%%%%%%%%%%%%%%%%%%%%%%%%%%%%%%
%%%%%%%%%%%%%%%%%%%%%%%%%%%%%%%%%%%%%%%%%%%%%%%%%%%
	
\section{Introduction} \label{sec:intro}

\paragraph*{}
Direct bandgap quantum dots (QDs) have long been studied for their potential use in quantum computing architectures \cite{laddQuantumComputers2010,mortonEmbracingQuantumLimit2011,awschalomQuantumSpintronicsEngineering2013}.  The deeper confining potential of self-assembled QDs allows for operations at higher temperatures compared to competing architectures such as gate defined dots or silicon dopants \cite{laddQuantumComputers2010}.  The combination of deep confining potentials and the opportunity to tune optical emission energies with QD size make III-V self-assembled QDs a particularly strong candidate for photonic quantum technologies \cite{bonadeoCoherentOpticalControl1998}. Indeed, epitaxially-grown self-assembled InAs QDs in GaAs have the best photonic performance of all currently available quantum emitters, with near-perfect radiative quantum efficiency, extremely high emission fraction into the zero-phonon line, and excellent coherence \cite{degreveUltrafastCoherentControl2011, stockillQuantumDotSpin2016, somaschiNearoptimalSinglephotonSources2016, atatureMaterialPlatformsSpinbased2018}.  Moreover, numerous experiments have already used InAs QDs to demonstrate quantum functionality such as entanglement generation or teleportation \cite{gaoQuantumTeleportationPropagating2013, stockillPhaseTunedEntangledState2017, delteilGenerationHeraldedEntanglement2016}.

\paragraph*{}
Qubits can be created when an InAs QD is charged with a single electron or hole whose up and down spin projections define the logical basis states of the qubit \cite{economouScalableQubitArchitecture2012, jenningsSelfAssembledInAsGaAs2020}. Such spin qubits can be controlled through the optical formation of trions, which are three-particle complexes consisting of an optically-generated exciton coupled to the additional resident charge. Optical selection rules for the trion transitions provide the means by which photons (``flying" qubits) initialize, manipulate, and readout the projection of the resident spin in order to implement quantum information functions \cite{economouScalableQubitArchitecture2012}. In this approach, the qubit lifetime is limited by the spin lifetime. Nuclear spins limit the electron spin coherence times \cite{merkulovElectronSpinRelaxation2002,khaetskiiSpinRelaxationSemiconductor2000,khaetskiiElectronSpinDecoherence2002}. In contrast, holes, which have weaker hyperfine interaction with the nuclei, can achieve spin coherence times that are an order of magnitude longer than that of electrons \cite{ebleHoleNuclearSpin2009,testelinHolespinDephasingTime2009} and may approach one second under certain conditions \cite{gillardFundamentalLimitsElectron2021}. An additional benefit of hole states is the presence of spin-orbit (SO) interactions that allow for control schemes not available with electrons \cite{kloeffelStrongSpinorbitInteraction2011,aresNatureTunableHole2013,crippaElectricalSpinDriving2018,watzingerGermaniumHoleSpin2018}.  
	
\paragraph*{}
Quantum Dot Molecules (QDMs) can be created when two QDs dots are stacked along the crystal growth axis and separated by a barrier thin enough to allow tunneling to create delocalized states with molecular orbital symmetry \cite{jenningsSelfAssembledInAsGaAs2020, dotyAntibondingGroundStates2009}.  In such QDMs, an electrical bias can be used to tune the states of the two QDs in and out of resonance with one another, controlling the formation of the molecular states \cite{bayerCouplingEntanglingQuantum2001,krennerDirectObservationControlled2005,ortnerControlVerticallyCoupled2005,stinaffOpticalSignaturesCoupled2006,dotyOpticalSpectraDoubly2008,dotyAntibondingGroundStates2009,dotyHolespinMixingInAs2010,liuSituTunableFactor2011}. For QDMs in which holes tunnel between the two QDs, important new effects can emerge in the presence of both a) spin-orbit coupling and b) symmetry breaking that arises from a lateral offset between the QDs that comprise the QDM. The interaction of spin-orbit coupling and symmetry breaking leads to delocalized molecular hole states composed of basis states with opposite hole spin character in each dot \cite{dotyHolespinMixingInAs2010}. This ``hole spin mixing" provides a tool for addressing a significant challenge for optically-controlled quantum information processing with InAs QDs. 
	
\paragraph*{}
In single InAs QDs it is not possible to optically drive a coherent rotation between the up and down spin projections (qubit basis states) without applying a transverse magnetic field. However, the application of a transverse magnetic field also introduces precession between the two qubit basis states, which prevents non-destructive readout.  The existence of hole spin mixing in QDMs provides a solution to this problem. Specifically, the spin projection of a hole in one of the QDs serves as the qubit basis. Coherent rotations between these two basis states can be achieved via a $ \Lambda$-transition that couples both spin basis states to a single delocalized molecule-like trion state. Importantly, such a trion state has allowed optical transitions to both qubit basis states because of hole spin mixing. Thus the presence of hole spin mixing makes it possible to initialize, manipulate, and readout hole-spin qubit states in QDMs without a transverse magnetic field, which in turn eliminates the problems with non-destructive readout \cite{economouScalableQubitArchitecture2012}. 
	
\paragraph*{}
The magnitude of hole spin mixing is critically important for quantum device operation \cite{economouScalableQubitArchitecture2012, vezvaeeAvoidingLeakageErrors2023}.  While hole spin mixing increases with the lateral offset between the two QDs \cite{climenteTheoryValencebandHoles2008,dotyHolespinMixingInAs2010,rajadellLargeHoleSpin2013,planellesSymmetryinducedHolespinMixing2015,maHoleSpinsInAs2016}, this offset is extremely difficult to control during growth. A lateral electric field, with a  gradient that is asymmetric between the two dots in the QDM, can mimic the effects of a lateral offset to enhance hole spin mixing \cite{maHoleSpinsInAs2016}, but the electrodes that would create such a field are extremely difficult to incorporate into photonic devices. We are evaluating the opportunity to enhance hole spin mixing in QDMs by replacing the \ce{GaAs} inter-dot barrier with a \ce{GaBi_xAs_{1-x}} alloy, without the need for a complicated electric field structure. The incorporation of Bi is expected to both lower the valence band edge (VBE) and increase spin-orbit interactions because Bi is heavier than As and has a strong spin-orbit coupling.
	
\paragraph*{}
In our previous work \cite{linIncorporationRandomAlloy2019}, we studied the energies of hole states in QDMs with \ce{GaBi_xAs_{1-x}} barriers at zero applied electric field (i.e.\ away from resonant coupling). In this paper, we study the hole states as a function of applied constant transverse electric field, which tunes the hole states of the two QDs through energetic resonance. In Section~\ref{sec:theory_and_methods}, we outline the atomistic tight-binding (TB) method that we use to calculate the wavefunctions of hole states at various electrical biases across the dots, as well as a schematic overview of tuning the dots states with an electric field.  In Section~\ref{sec:resonance}, we show that a maximum three-fold increase of tunnel coupling is achieved via alloying.  Additionally, we have a discussion on local potentials created in the barrier region by the random nature of the alloy distribution.  In Section~\ref{subsec:geo-effects}, we analyze the separate contributions of the additional strain and orbital fluctuations introduced by the alloying.  We show that the strain introduced by alloying can shift the field strength required to achieve resonance by applying additional asymmetry to the system.  However, when applied in isolation without the orbital effects of the alloy, the alloy strain does not affect the intrinsic tunneling behavior of the hole states.  Orbital changes introduced by the alloying, on the other hand, do not affect the location of the resonance and better show the increased tunnel coupling from lowering the barrier energy. Importantly, the combination of alloy strain and orbital effects lower the barrier height significantly more than either effect individually, which results in the significant increase in tunnel coupling strength. Finally, in Section~\ref{subsec:geo-spacing}, we have a brief discussion on the optimization of alloy effects by varying the dot-to-dot separation.  In a forthcoming paper we will build on these results to probe the extent to which the incorporation of Bi increases hole spin mixing or g-factor tunability in QDMs, furthering the opportunities to use these QDMs for quantum technologies.

%%%%%%%%%%%%%%%%%%%%%%%%%%%%%%%%%%%%%%%%%%%%%%%%%%%
%%%%%%%%%%%%%%%%%%%%%%%%%%%%%%%%%%%%%%%%%%%%%%%%%%%

\section{Theory and Motivation} \label{sec:theory_and_methods}

%%%%%%%%%%%%%%%%%%%%%%%%%%

\subsection{Atomistic tight-binding model} \label{subsec:theory-tb}

\begin{figure}[ht]
	\centering
	\subfloat[Aligned, fully alloyed barrier]{
		\includegraphics[width=0.23\textwidth]{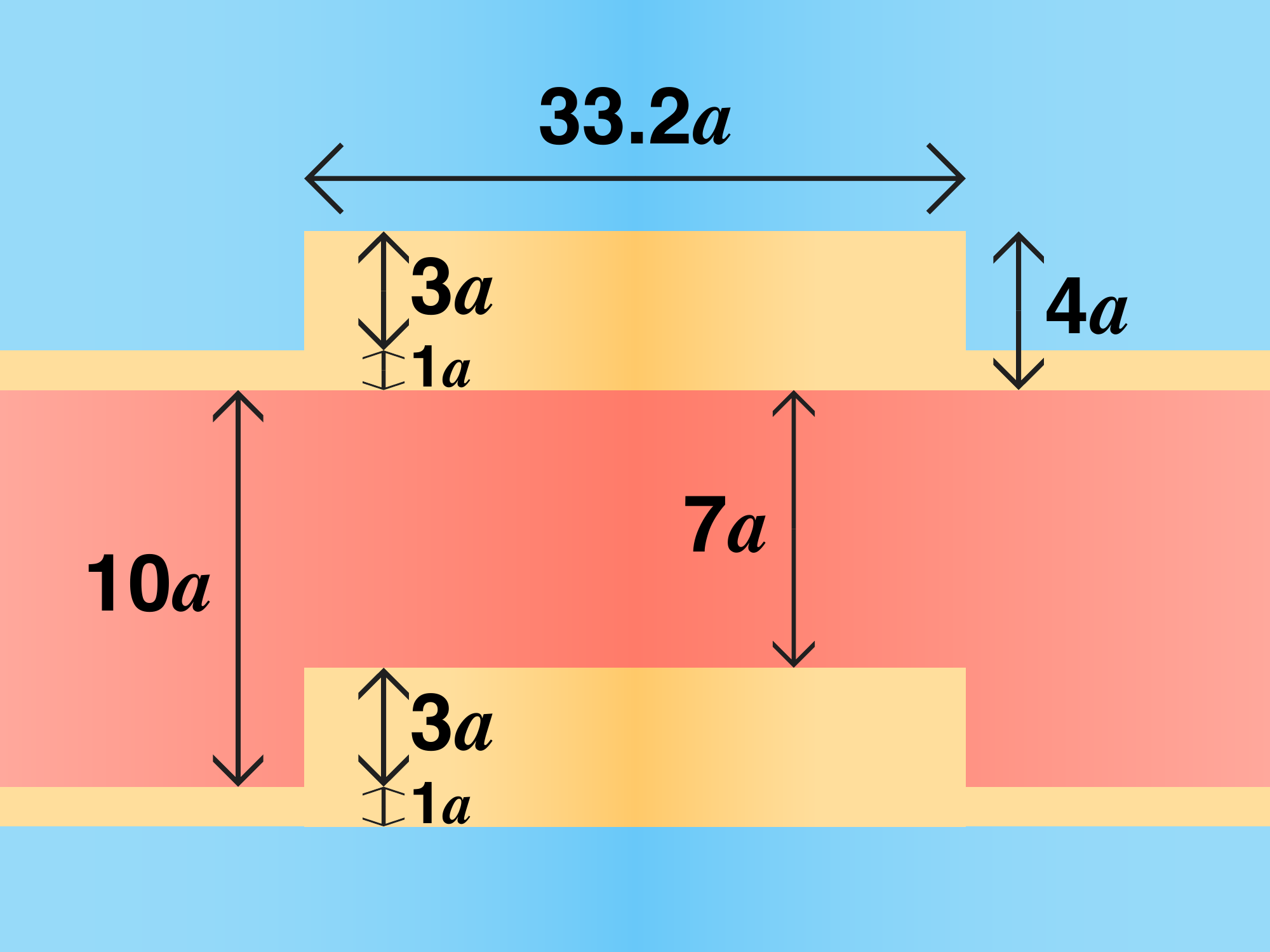}
		\label{fig-sub:system_geo-full}
	}
	\subfloat[Aligned, partially alloyed barrier]{
		\includegraphics[width=0.23\textwidth]{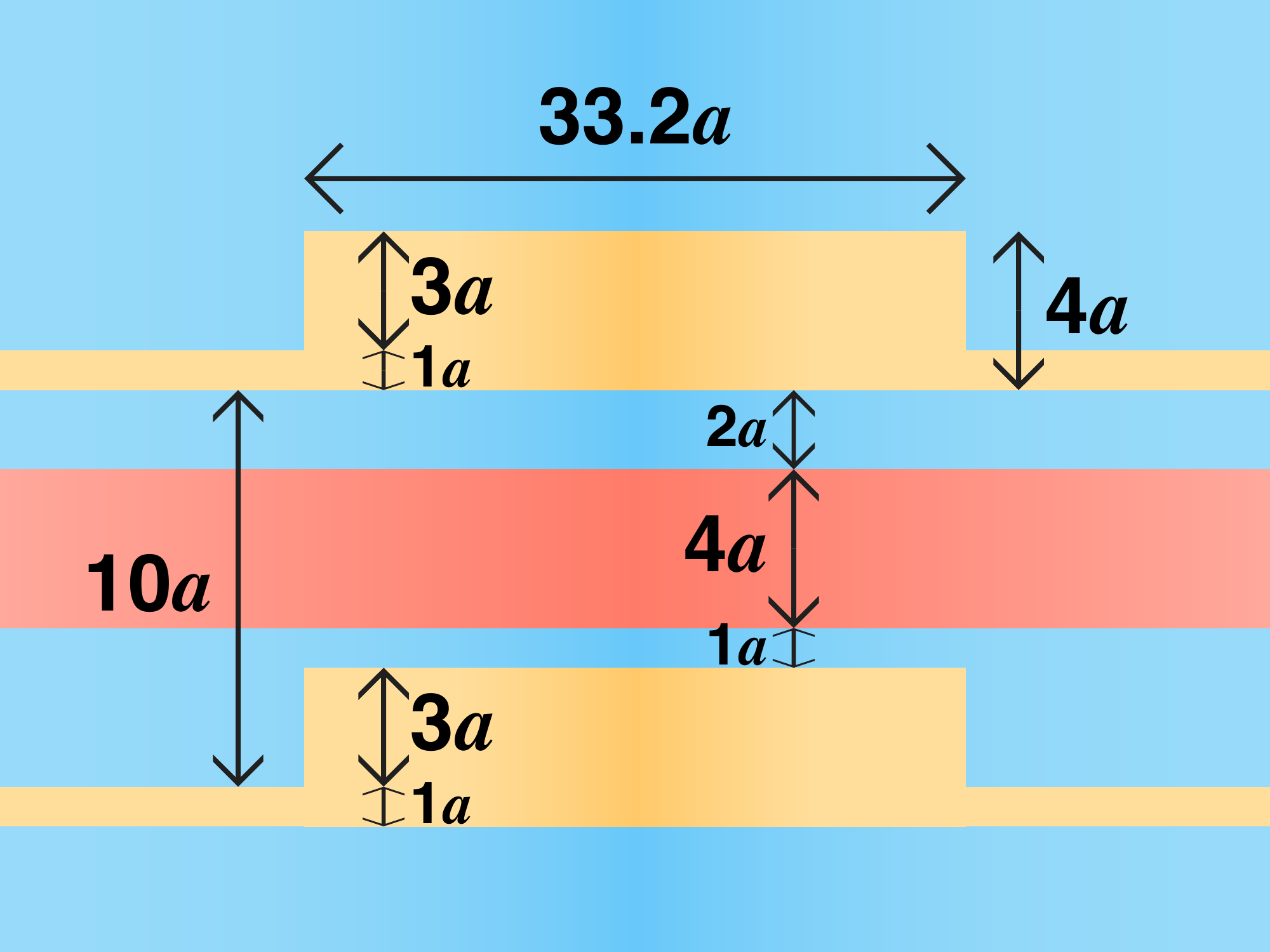}
		\label{fig-sub:system_geo-part}
	}
	\caption[General geometry of \ce{InAs} QDM, with proposed \ce{GaBiAs} in barrier region.]{\small Schematic representation of system geometry, same as Ref~\cite{linIncorporationRandomAlloy2019}: aligned dots with (a) a fully alloyed barrier of \ce{GaBiAs} spanning the full inter-dot region or (b) a partially alloyed barrier with a layer of \ce{GaBiAs} embedded in between \ce{GaAs}.  \ce{GaAs} in blue; \ce{InAs} in yellow; \ce{GaBiAs} in red.  \newline $a$ represents lattice constants.  $x$ and $z$ axes not to scale.}
	\label{fig:system_geo}
\end{figure}

\paragraph*{}
Our model for the QDM system is built upon the same atomistic $ sp^3s^* $ nearest-neighbor tight-binding (TB) model outlined in our previous paper \cite{linIncorporationRandomAlloy2019}.  As shown in Figure~\ref{fig:system_geo}, two vertically stacked QDs, which are grown experimentally by molecular-beam epitaxial deposition of \ce{InAs} on \ce{GaAs}, form the trapping potential of the QDM.  The two \ce{InAs} dots are separated by a \ce{GaAs} or \ce{GaBiAs} barrier.  The layer structure and alloy composition of the barrier are tunable during growth.  During growth, the strain induced by the dot in the first \ce{InAs} layer acts as a nucleation site for the growth of the second dot \cite{wasilewskiSizeShapeEngineering1999}.  Thus, the two dots of the QDM are naturally stacked in the growth direction.

\paragraph*{}
The atomistic TB model has been well studied and verified for semiconductor heterostructures \cite{harrisonBondOrbitalModelProperties1973,harrisonBondorbitalModelII1974,pantelidesStructureValenceBands1975,voglSemiempiricalTightbindingTheory1983,grafElectromagneticFieldsDielectric1995,maHoleSpinsInAs2016}.  Previously, in Ref~\cite{linIncorporationRandomAlloy2019} we have validated the efficacy of using an atomistic TB model to calculate the valence band energies of bulk random alloy \ce{GaBi_xAs_{1-x}} and \ce{InAs}/\ce{Ga(Bi)As} QDMs.  The alloy TB Hamiltonian was constructed atomisticly using the TB parameters of the material at each atomic site.  This not only allows us to study random alloy distributions, but is also necessary in capturing the unique alloy behavior of  \ce{GaBi_xAs_{1-x}} that is lost if parameters were averaged over the alloy concentration (such as in a virtual crystal approximation) \cite{linIncorporationRandomAlloy2019}.  The TB Hamiltonian is then modified to account for lattice mismatch induced strain by scaling the TB parameters in accordance with the relaxed atomic positions calculated with a valence force field (VFF) model.  Wavefunctions of interest are found by iterative diagonalization of the Hamiltonian, done with the ARPACK implementation of the Arnoldi method. 

\paragraph*{}
In this paper, we extend the model of our previous paper to include an electric field.  This simulates the application of an electrical bias during the operation of QDM devices.  An electric field is added to the TB Hamiltonian by applying the scalar potential, $ \Phi(\vec{R}_n) $, to the on-site TB parameters:
\begin{equation}
	\epsilon_{\alpha,\vec{R}_n} \equiv \epsilon^0_{\alpha,\vec{R}_n} - e \Phi(\vec{R}_n).
\end{equation}
$ \epsilon^0_{\alpha,\vec{R}_n} $ is the zero-field on-site energy of the orbital $\alpha$ at atomic site $\vec{R}_n$, and $e$ is the electron charge.  The scalar potential at each atomic site is stored as a separate file, and added to the TB Hamiltonian during the diagonalization process.

\paragraph*{}
For the operation of the QDMs, the electric bias is applied along the stacking axis of the two dots (the $z$-direction).  This creates a constant electric field, which has the scalar potential
\begin{equation}
	\Phi(\vec{R}_n) = -E z.
\end{equation}
The scalar potential is set to zero at $z=0$, which is chosen to be at the bottom of the lower dot. As a result, the energy of states confined in the bottom dot scale minimally with the applied electric field; and the energy of states confined in the top dot scale linearly with the applied electric field.  The visually distinct slopes aids us in identifying the states when plotting energy versus applied field.

\paragraph*{}
All calculations in this paper are done for cylindrical-disk dots, with a radius of $1 6.6a $ ($a$ being the lattice constant), a height of $ 3a $ (not including the wetting layer), and a $ 1a $ thick wetting layer, as shown in Fig.~\ref{fig:system_geo}.  Dot-to-dot separation is $10a$, unless otherwise noted.  A computational box of size $ 144a \times 144a \times 45a $ is used for calculating the strain-relaxed atomic positions with the VFF parameters.  With the new atomic positions, a $ 50a \times 50a \times 34a $ cutout of the relaxed lattice containing the QDM is made.  Our TB Hamiltonian is constructed for the smaller box saving significant computation time during diagonalization.

\paragraph*{}
In this paper, only results for the perfectly aligned dots are shown.  Calculations for offset dots were also performed, showing qualitatively very similar results supporting the same conclusions.  We chose to omit results for the offset dots for brevity, as the offset does not affect the spin-preserving tunnel coupling of the hole states.  The lateral offset between the two dots is important for the creation of spin-mixed states between the two dots, which is outside the scope of this paper.  Additionally, the majority of results will be shown for the fully alloyed barrier, representing the maximum effect of the \ce{Bi} alloy.  In the latter half of this paper, we shall also present results for the partially alloyed barrier, to demonstrate alloy strain effect on the tunnel coupling of hole states.

%%%%%%%%%%%%%%%%%%%%%%%%%%

\subsection{Tunnel resonant hole states} \label{subsec:theory-tunnel}

\begin{figure}[ht]
	\centering
	\subfloat[Dots electrically tuned to resonance]{
		\includegraphics[width=0.4\textwidth]{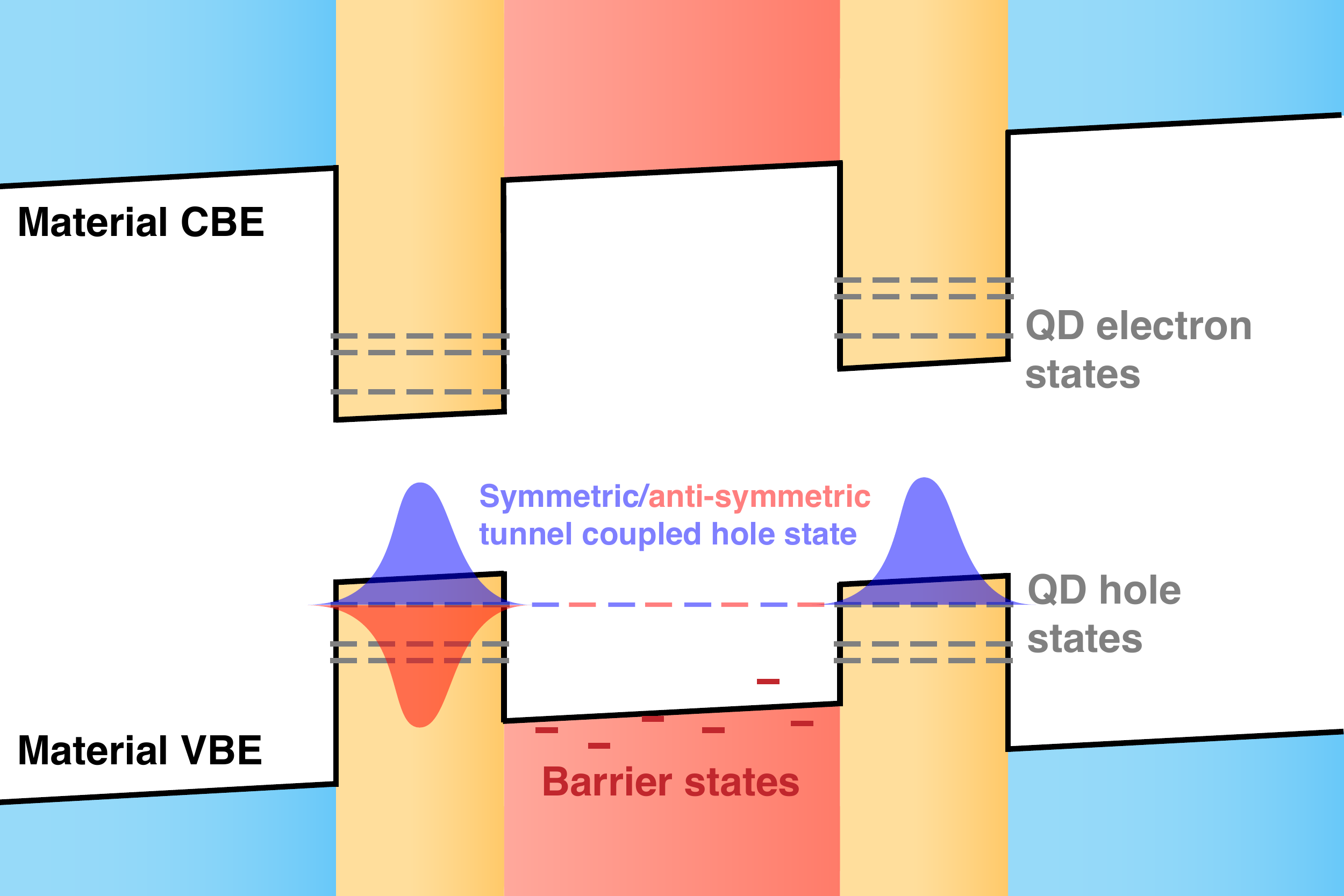}
		\label{fig-sub:system_band_edge-resonance}
	}
	\hspace{0em}
	\subfloat[Zero electrical bias]{
		\includegraphics[width=0.22\textwidth]{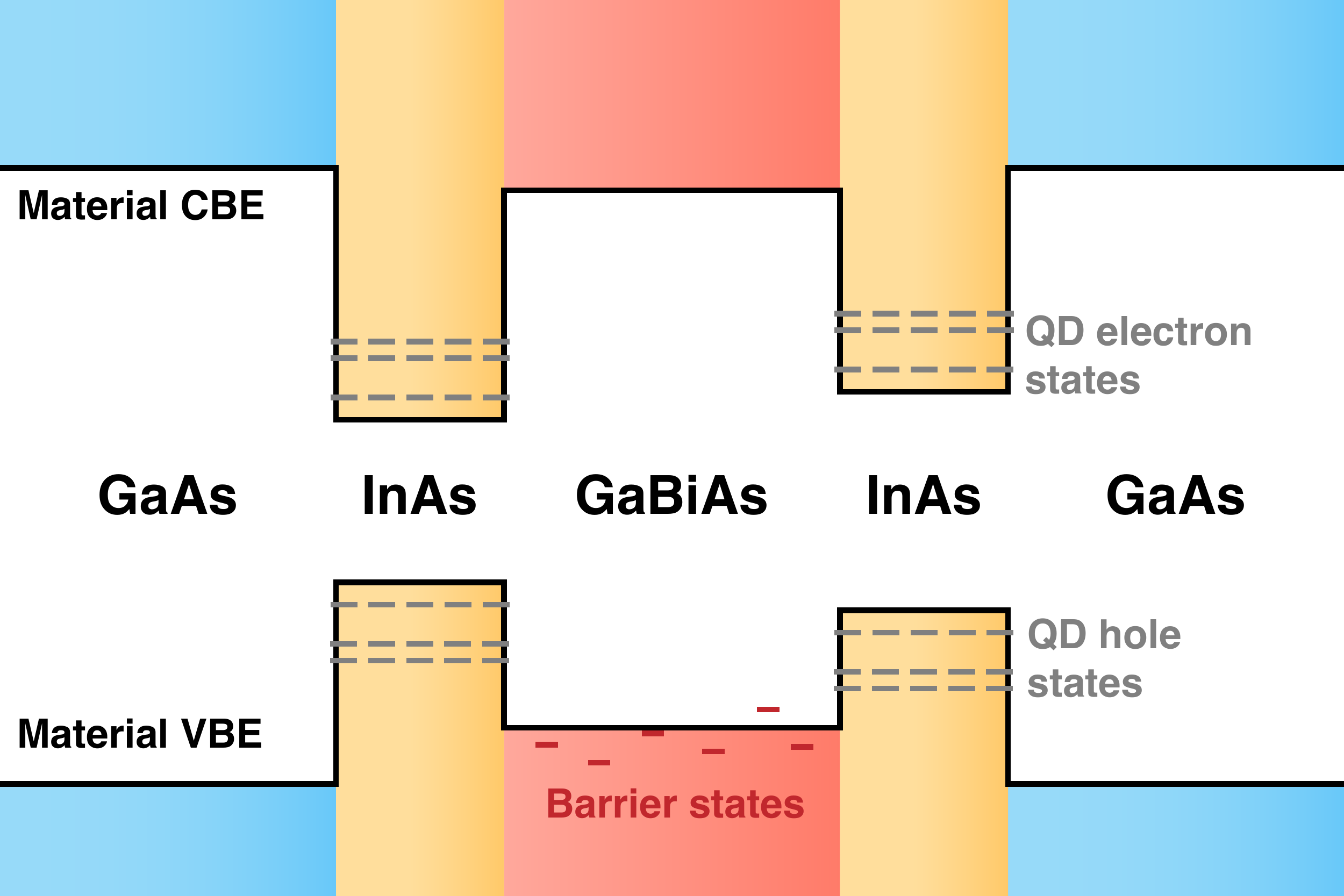}
		\label{fig-sub:system_band_edge-zero_field}
	}
	\subfloat[Electrical bias applied past resonance]{
		\includegraphics[width=0.22\textwidth]{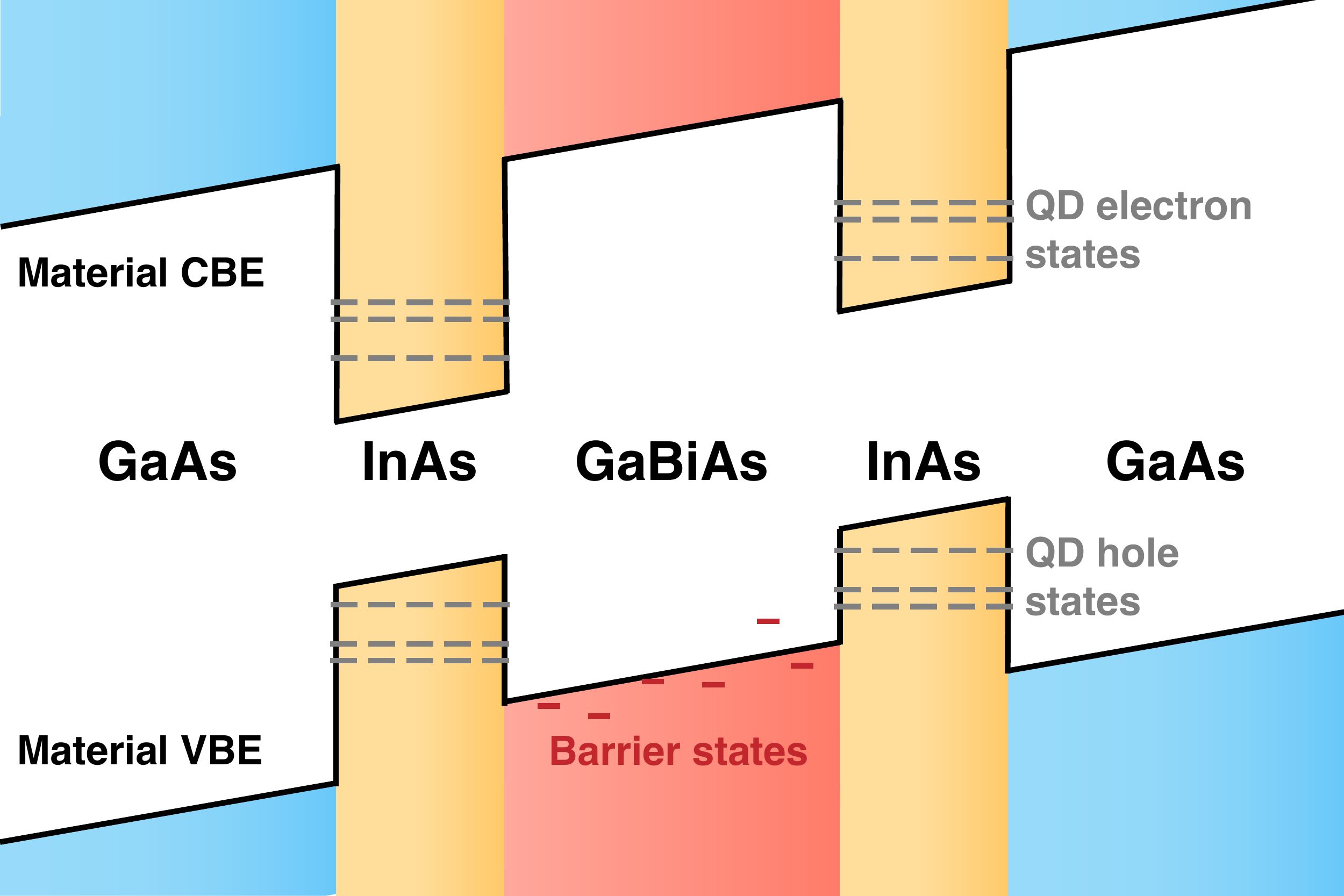}
		\label{fig-sub:system_band_edge-with_field}
	}
	\caption[Schematic of applying an electrical bias to a \ce{InAs} QDM with \ce{GaBiAs} in barrier.]{\small Band diagram of InAs/GaBiAs QDMs electrically tuned with a field in the stacking direction of the dots.  (a) At resonance, hole states tunnel couple across the two dots to from symmetric and anti-symmetric molecular states. (b) At zero electrical bias, the hole states are well separated, energetically, by the difference in strain experienced by each dot.  (c) Tuned past resonance, higher energy hole states and states localized to the alloyed barrier region, can be brought to the same energy region as the ground state of the lower dot.}
	\label{fig:system_band_edge}
\end{figure}

\paragraph*{}
Away from resonance (Figure~\ref{fig-sub:system_band_edge-zero_field} and Figure~\ref{fig-sub:system_band_edge-with_field}), the hole states are well confined to their respective dots.  The zero-field energies of the dot states are different for the two dots due to the difference in strain caused by the placement of the wetting layer, breaking mirror symmetry.  The tunnel coupling strength between the states of opposing dots can be characterized by the hybridization of the states when the dots are brought into energetic resonance (Figure~\ref{fig-sub:system_band_edge-resonance}).  We investigate the coupling strength by looking at the behavior of the hole state energies as the holes are brought into resonance by a uniform electric field applied along the stacking axis of the two QDs \cite{brackerEngineeringElectronHole2006,dotyAntibondingGroundStates2009}.  The interaction between the states hybridizes the states to form symmetric and anti-symmetric pairs, i.e., molecular states.  The anti-crossing energy between the symmetric and anti-symmetric pair at resonance is representative of the strength of the tunnel coupling between the dots.

\paragraph*{}
The introduction of \ce{GaBiAs} in place of GaAs enhances the tunnel coupling in two distinct ways.  First, raising the valence band energy (lowering the energy barrier) of the barrier material weakens the tunnel barrier and promotes dot-to-dot tunnel coupling (Section~\ref{subsec:resonance-gabias_x}).  Second, since \ce{Bi} atoms are larger than \ce{As}, introducing \ce{Bi} to the \ce{GaAs} barrier introduces additional strain into the system.  If the amount of strain applied to the two dots by the alloy is not equivalent, the hole state energies of the two dot are shifted by different amounts, changing the potential bias at which resonance occurs.  Yet, the effect of alloy
strain alone does not change the strength of the tunnel coupling between the two dots (Section~\ref{subsec:geo-effects}).
	
\paragraph*{}
The random distribution of Bi atoms also creates pockets of local potential in the barrier, resulting in the presence of hole states with charge densities localized around a few atoms in the barrier (Section~\ref{subsec:resonance-gabias_config} and supplementary file).  At lower alloy concentrations, these states remain energetically separated from the well behaved dot states.  However, at higher alloy concentrations, the barrier VBE and these localized states in the barrier are brought into, or even above, the valence energy region of the dot states.  At a threshold concentration, these barrier states can be tuned in and out of energetic resonance with the dot states by an electric field (Figure~\ref{fig-sub:system_band_edge-with_field}).  Generally, the barrier states only very weakly couple to the dot states, but the strength of coupling is very dependent on the strength of the local potential and how close, spatially, the barrier state is localized with respect to the dot.
	
%%%%%%%%%%%%%%%%%%%%%%%%%%%%%%%%%%%%%%%%%%%%%%%%%%%
%%%%%%%%%%%%%%%%%%%%%%%%%%%%%%%%%%%%%%%%%%%%%%%%%%%
	
\section{Coherent Tunneling of Hole States Through an Alloyed Barrier} \label{sec:resonance}

%%%%%%%%%%%%%%%%%%%%%%%%%%

\subsection{Concentration dependence} \label{subsec:resonance-gabias_x}

\paragraph*{}
Figure~\ref{fig:spacing-7_e-field_z1.0d4} shows the energy levels of the hole states in response to an electric field applied along the growth axis. The hole state energies are shown as the energy of the corresponding valence states, such that the lowest energy hole state is at the top of the band of valence states.  We have set the potential to be zero at the bottom of lower dot, so that the energies of the bottom dot states remain unchanged when the field is applied, while the energies of states mostly localized to the top dot increase linearly with the field strength and dot separation.  The state with the lowest hole energy for each of the two dots, including the hybrid states they form, are colored black; higher energy states are colored blue; states that have a significant wavefunction density in the barrier region between the dots are colored red.  The coloring of the state are determined with calculations of the probability distribution of the wavefunction (Appendix~\ref{appx:rho}) and spatial plots of the charge density (see the supplementary file).

\paragraph*{}
The states of most interest are the ground state and the first excited state, or the lowest energy hole state for each of the dots (black in Figure~\ref{fig:spacing-7_e-field_z1.0d4}).  Looking at the energy difference between the ground and excited states at each electric field strength, the minimum energy difference defines the location and strength of the anticrossing.  The larger the anticrossing energy, the stronger the coupling between the two states, making it easier to tunnel from one dot to the other.  Up to 6 \% Bi, we see a steady increase of anticrossing energy between the two lowest energy hole state of the respective top and bottom dots.  We also see that the resonance occurs at progressively larger field strengths.  The change in resonance location is not due to the tunneling behavior.  Rather, the shift results from the increasing initial (zero-field) difference between the ground and first excited states that is due to the increased strain on the top dot introduced by the alloy \cite{linIncorporationRandomAlloy2019}.  The presence of \ce{Bi} surrounding the bottom dot adds to the asymmetry of the QDM, causing the larger energy difference between the top and bottom dot states. The larger initial difference between the states in the top and bottom dots does not result in a larger anticrossing energy, just an increase in the field strength needed to bring the states into resonance.  The increase in anticrossing energy results from the stronger tunnel coupling between the two dots, regardless of where the zero-field energies lie relative to each other.  We shall revisit this in Section~\ref{subsec:geo-effects}, where we will investigate the strain effects due to alloying.
	
\paragraph*{}
As the barrier \ce{GaBi_{x}As_{1-x}} band edge approaches the energy range of the dot levels, the dot states lose confinement and leak into the barrier \cite{linIncorporationRandomAlloy2019}.  As a consequence, starting at around 8\% Bi, where the GaBiAs barrier has similar valence band energies to the InAs dots, we see a loss in well-defined resonance behavior for the dot states.  Figures~\ref{fig-sub:spacing-7_e-field_0.08_z1.0d4} and \ref{fig-sub:spacing-7_e-field_0.10_z1.0d4} are thus qualitatively different from lower alloy concentrations.  The states drawn in red have significant charge density in the barrier and show the encroachment of the \ce{GaBiAs} valence band energies on the \ce{InAs} dot states.

\paragraph*{}
Appendix~\ref{appx:rho} discusses the increase of charge density in the barrier region as alloy concentration is increased.  At 8\% Bi, not only is a large part of the wavefunction in the barrier, but the wavefunctions also start losing the nodal structure of dot confinement.  At 10\% Bi, the wavefunctions are almost entirely in the barrier, localized around few-atom Bi clusters, resulting in states strongly dependent on the alloy configuration.  In addition to what the wavefunction plots display, the leakage of charge into the barrier and the configurational variances of the alloy can also be seen by applying higher electrical biases, as we shall discuss in the next sub-section.

\begin{widetext}
	\onecolumngrid
	\begin{figure}[h!]
		\subfloat[0\% Bi]{
			\includegraphics[width=0.3\textwidth]{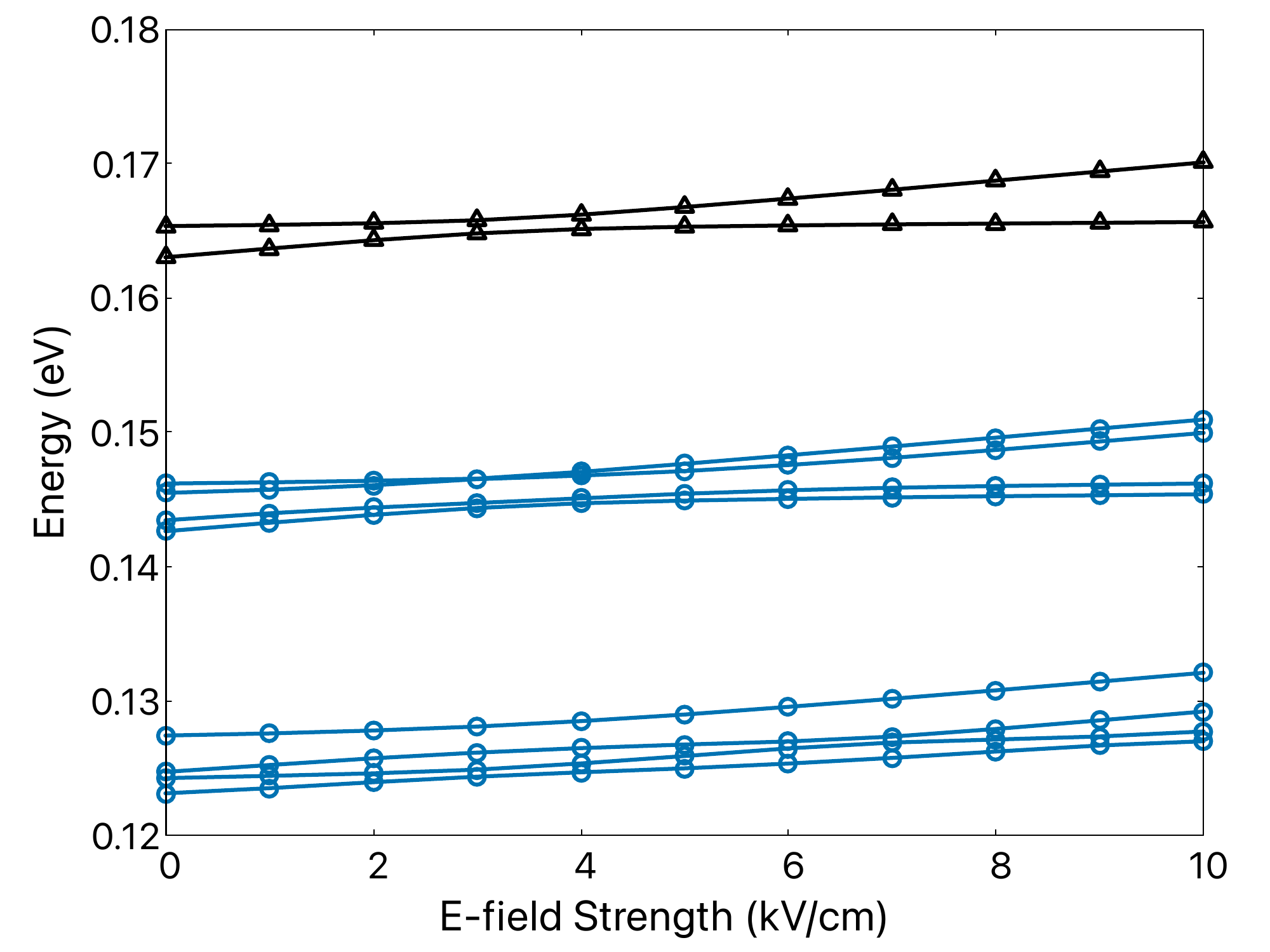}
			\label{fig-sub:spacing-7_e-field_0.00_z1.0d4}
		}
		\subfloat[2\% Bi]{
			\includegraphics[width=0.3\textwidth]{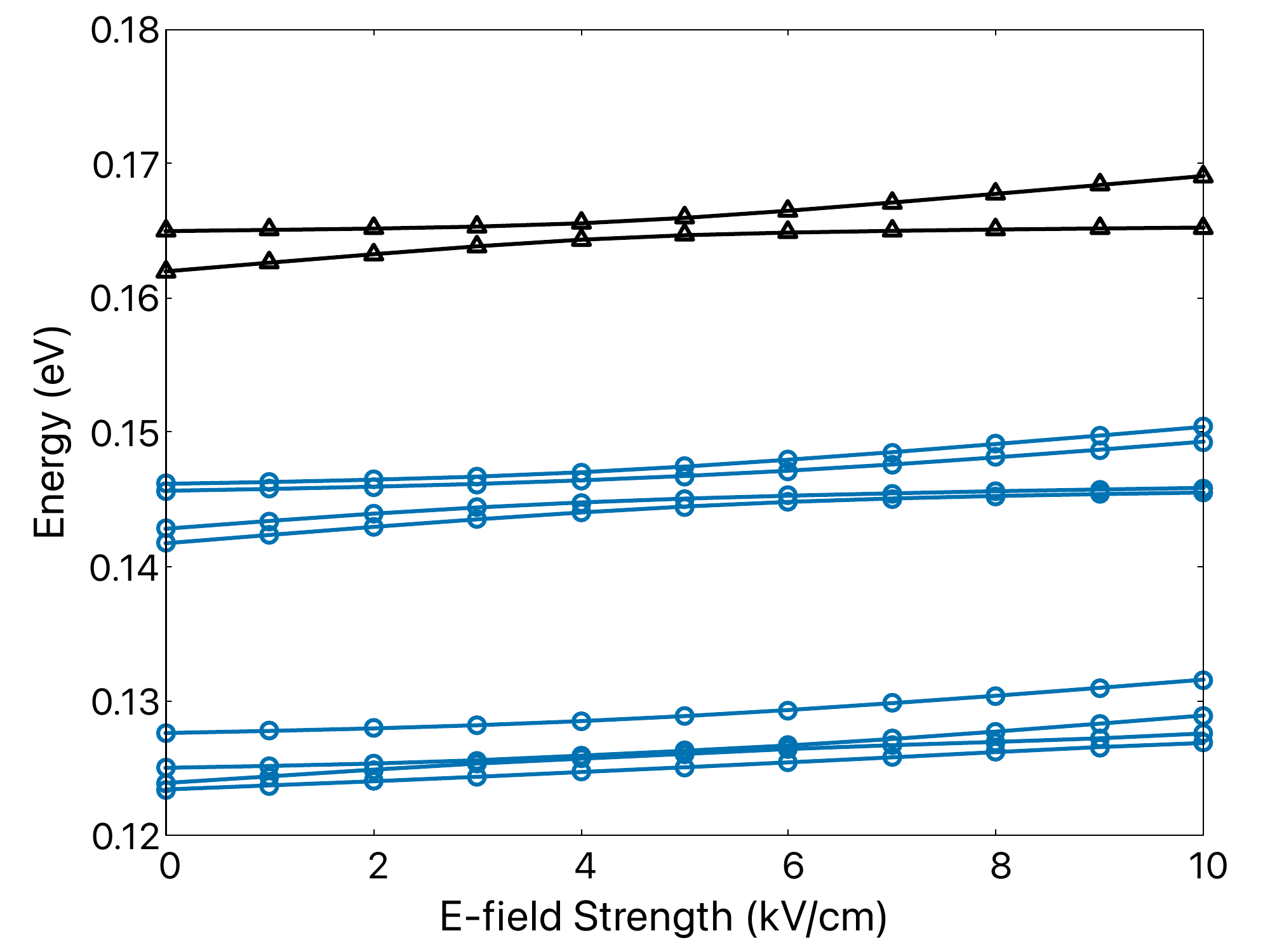}
			\label{fig-sub:spacing-7_e-field_0.02_z1.0d4}
		}
		\subfloat[4\% Bi]{
			\includegraphics[width=0.3\textwidth]{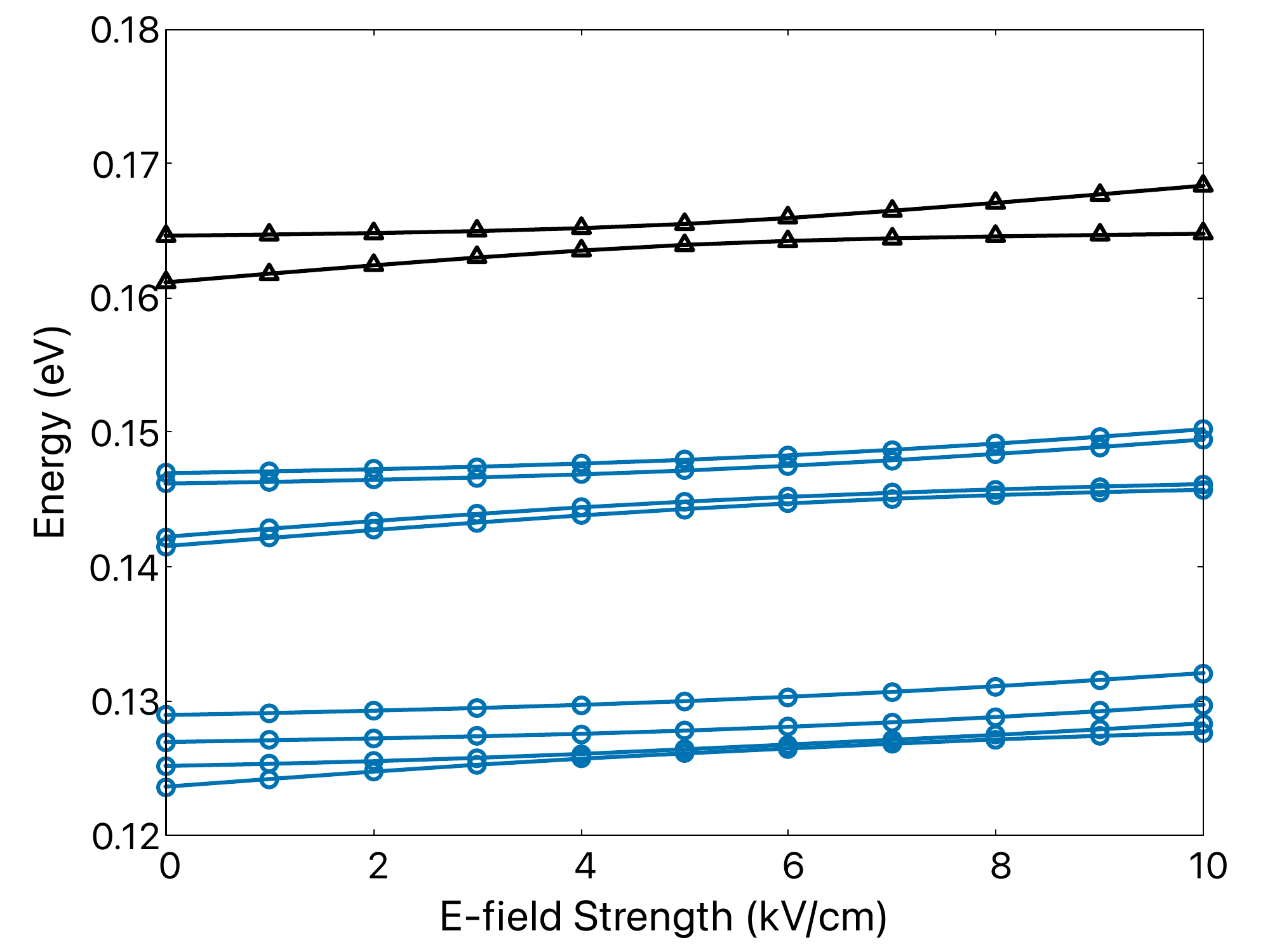}
			\label{fig-sub:spacing-7_e-field_0.04_z1.0d4}
		}
		\hspace{0em}
		\subfloat[6\% Bi]{
			\includegraphics[width=0.3\textwidth]{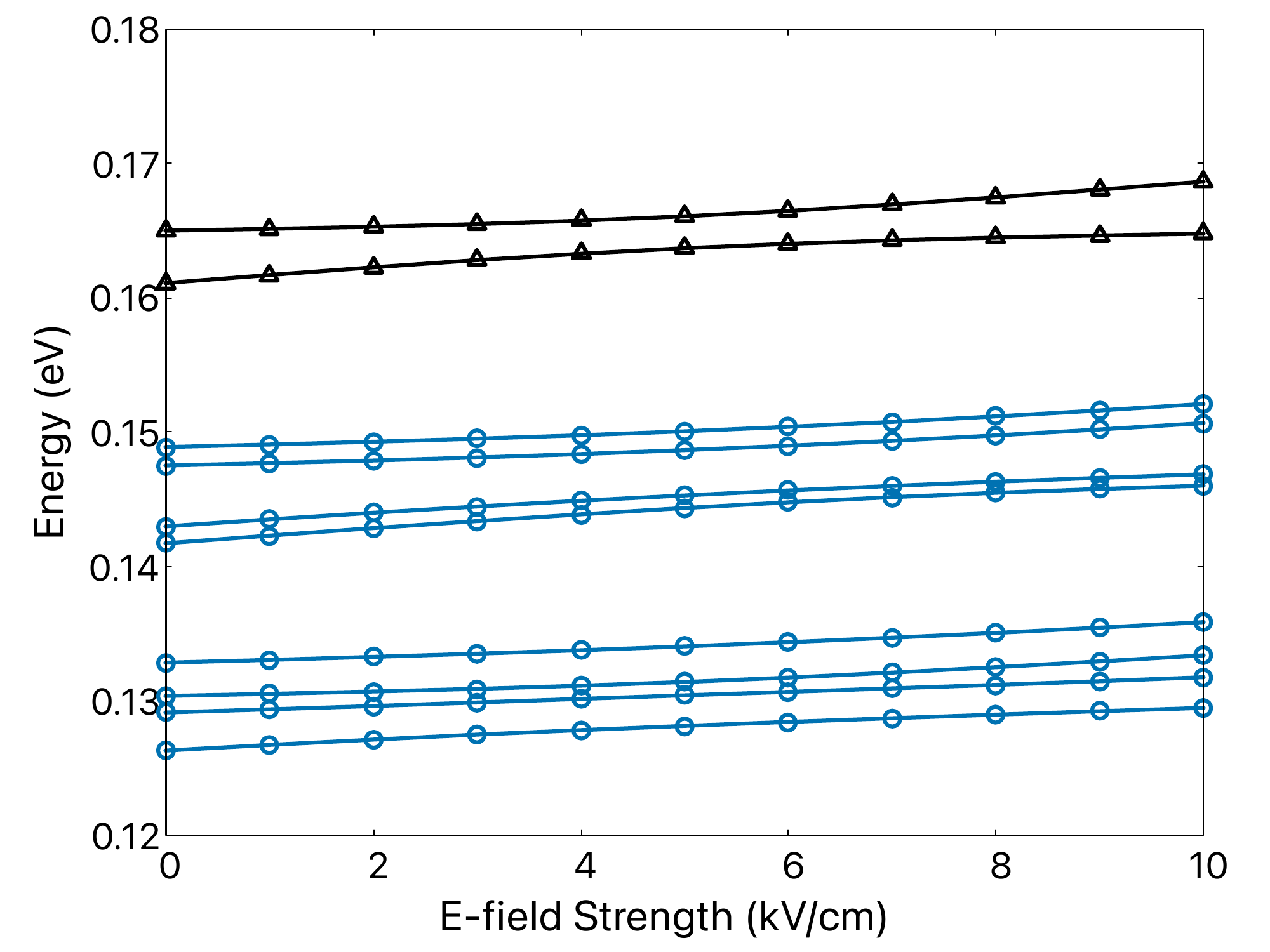}
			\label{fig-sub:spacing-7_e-field_0.06_z1.0d4}
		}
		\subfloat[8 \%]{
			\includegraphics[width=0.3\textwidth]{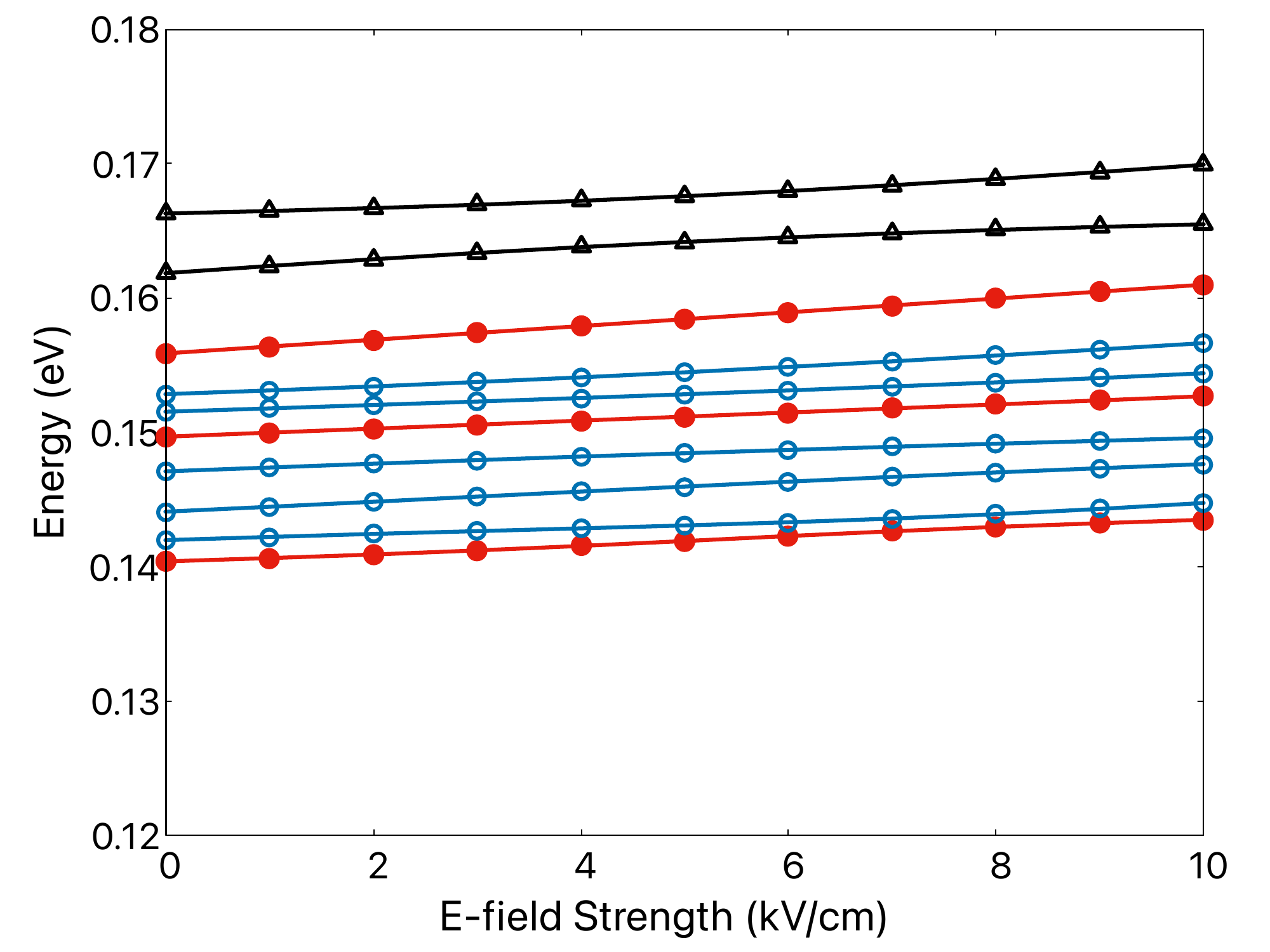}
			\label{fig-sub:spacing-7_e-field_0.08_z1.0d4}
		}
		\subfloat[10 \%]{
			\includegraphics[width=0.3\textwidth]{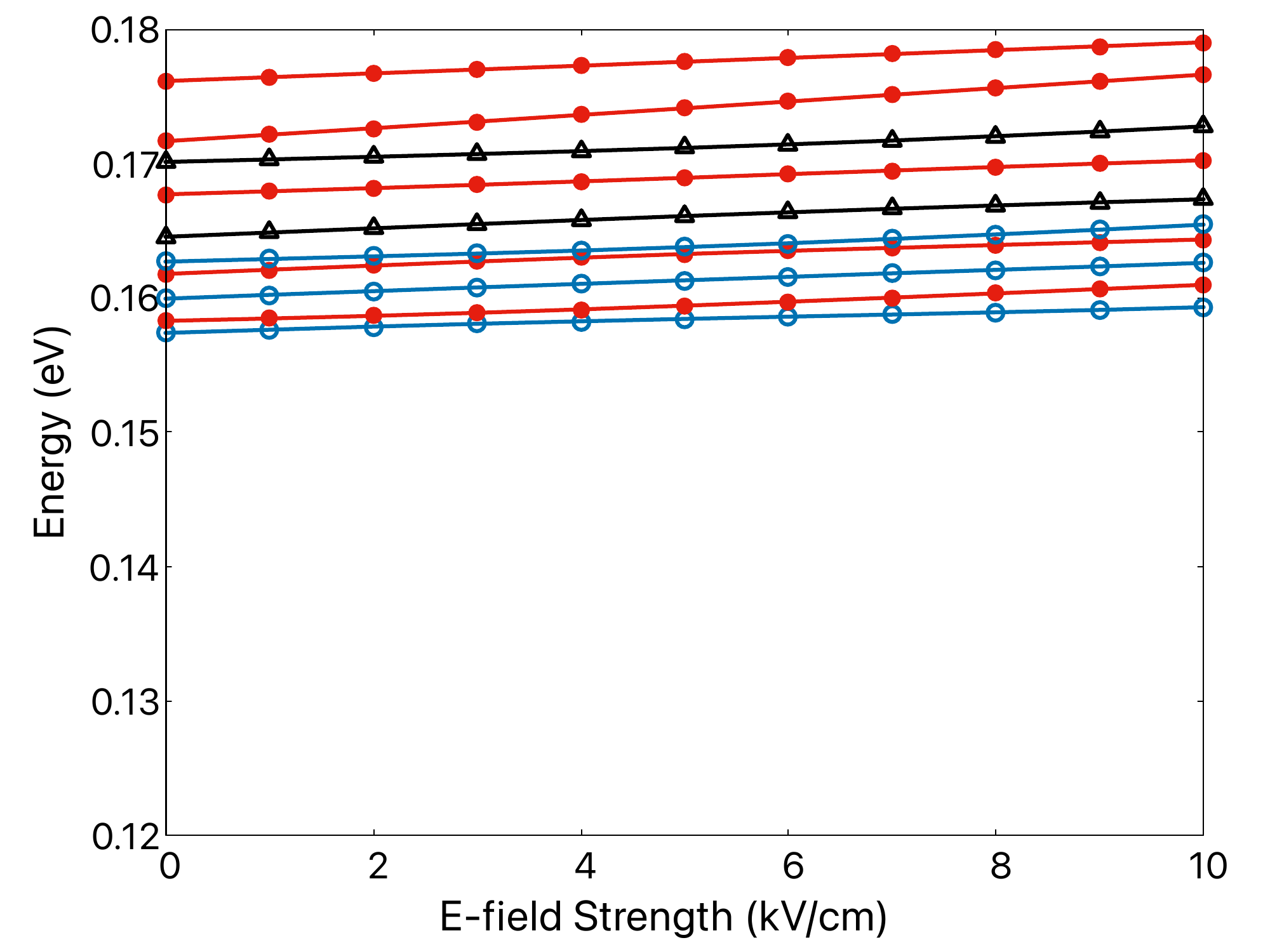}
			\label{fig-sub:spacing-7_e-field_0.10_z1.0d4}
		}
		\caption[Energy levels of an \ce{InAs} QDM with a \ce{GaBi_xAs_{1-x}} barrier,  under electric field up to\SI{10}{\kilo\volt\per\centi\meter}.]{\small Energy levels of an \ce{InAs} QDM with a \ce{GaBi_xAs_{1-x}} full barrier, where (a) $x=0.00$, (b) $x=0.02$, (c) $x=0.04$, (d) $x=0.06$,  (e)$x=0.08$, and (f) $x=0.10$. An electric field of up to \SI{10}{\kilo\volt\per\centi\meter} is applied along the lattice growth axis.  Black: the lowest energy hole state confined to either dot, and the molecular states they form.  Blue: higher energy hole states that are still well confined to the dots.  Red: states with charge densities highly localized to the barrier region.}
		\label{fig:spacing-7_e-field_z1.0d4}
	\end{figure}
\end{widetext}

\begin{figure}[ht]
	\includegraphics[width=0.45\textwidth]{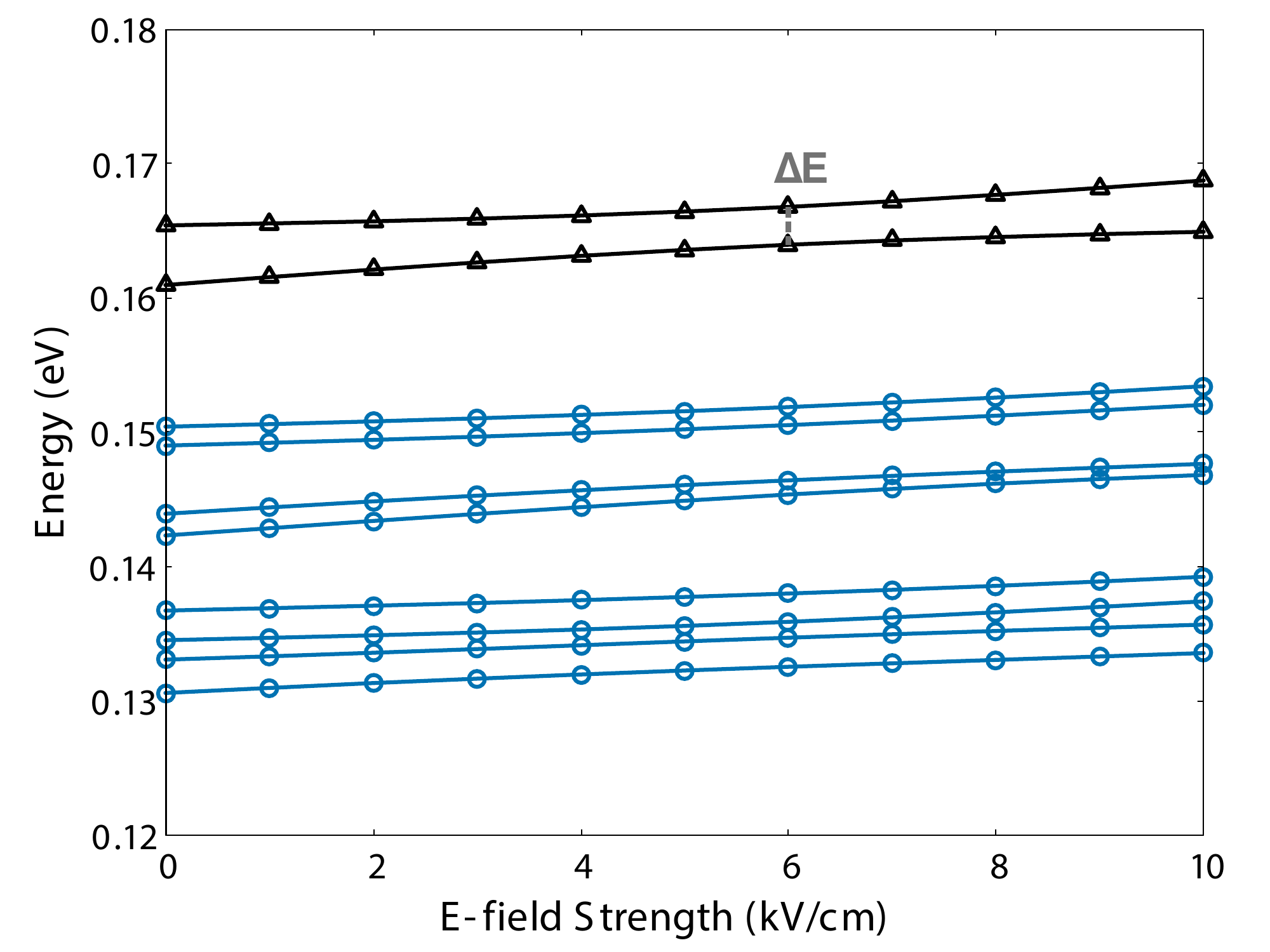}
	\caption[Energy levels of an \ce{InAs} QDM with a \ce{GaBi_{0.07}As_{0.93}} barrier, under electric field up \SI{10}{\kilo\volt\per\centi\meter}.]{\small Energy levels of an \ce{InAs} QDM with a \ce{GaBi_{0.07}As_{0.93}}, under an electric field up to \SI{10}{\kilo\volt\per\centi\meter} along the lattice growth axis.}
	\label{fig:spacing-7_e-field_0.07}
\end{figure}

\paragraph*{}
For the geometry of the dots given, we have found 7\% alloy to be the optimal concentration for tunnel coupling enhancement.  Figure~\ref{fig:spacing-7_e-field_0.07} shows the results for a QDM with a \ce{GaBi_{0.07}As_{0.93}} inter-dot barrier. A threefold increase in anticrossing energy from the non-alloyed case occurs.  A 7\% Bi alloy was the highest concentration we found where all the states were well confined in the dot.  While higher alloy concentrations can have larger anti-crossing energies, the states tend to lose their dot like characteristics. Controlled amount of leakage into the barrier can be beneficial in increasing the anticrossing energy and creating a more strongly coupled molecular state.  However, the large amount of confinement loss at 8\% and higher alloy concentrations leads to poorly behaved states that are more sensitive to the alloy disorder.
	
\begin{figure}[htb]
	\includegraphics[width=0.45\textwidth]{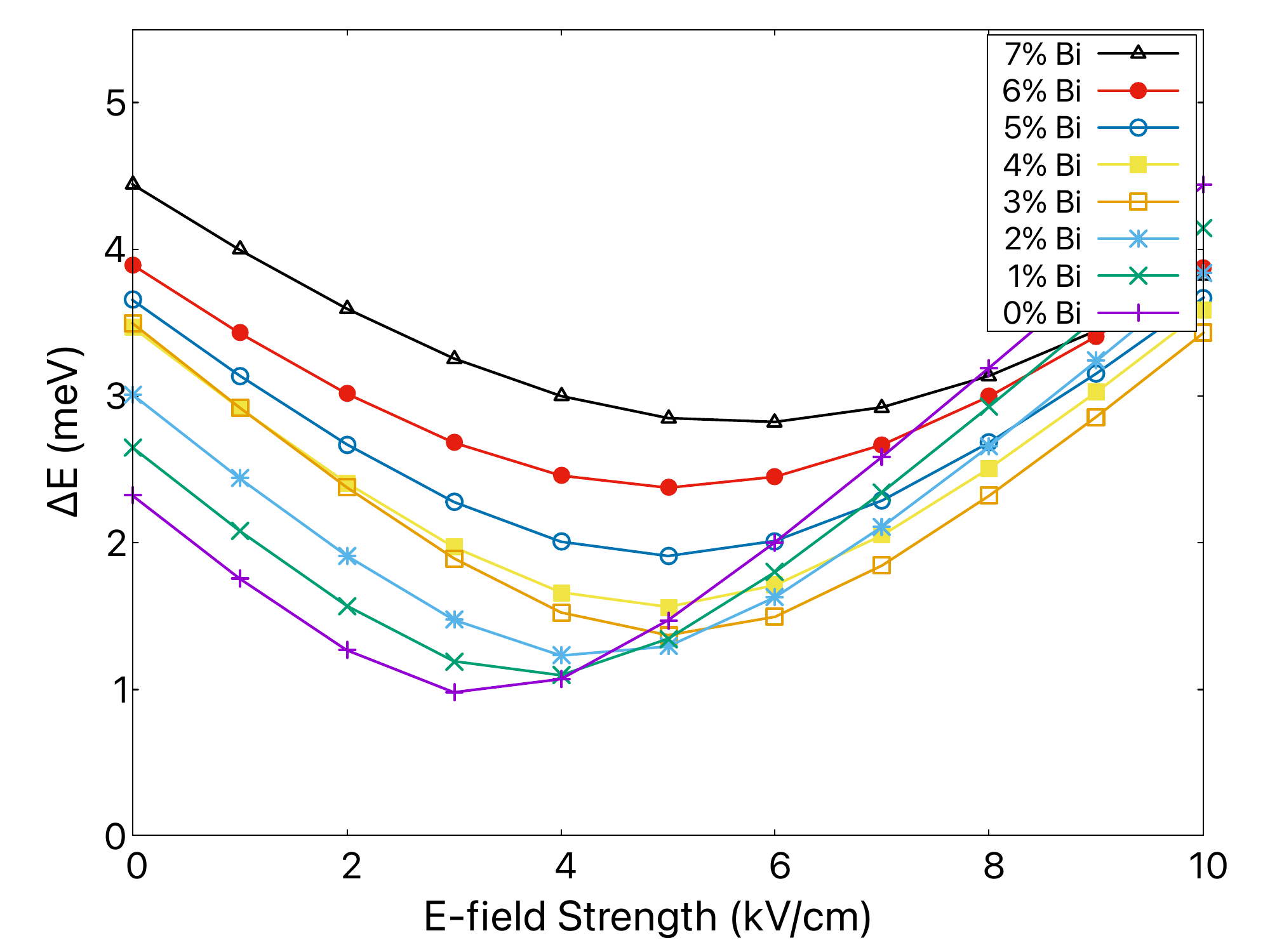}
	\caption[Anti-crossing energies for top two valence states in \ce{InAs} QDM.]{\small Energy difference between the top two valence states of a \ce{InAs} QDM with a full \ce{GaBi_{x}As_{1-x}} inter-dot barrier, obtained from a random alloy model.}
	\label{fig:crossing}
\end{figure}

\paragraph*{}
Figure~\ref{fig:crossing} shows the energy splitting between two lowest energy hole states as a function of the applied field for concentrations up to 7\%.  These two states represent the lowest energy state for the bottom dot and the top dot, or, in the case of resonance, the symmetric and anti-symmetric between the two dots.  The minimum energy difference occurs at tunnel resonance. The magnitude of the  difference at resonance is the anti-crossing energy and indicates the strength of the tunnel coupling.  From Figure~\ref{fig:crossing}, the resonance shift to higher electrical bias as alloy concentration increases is clear. Moreover, as the alloy concentration increases, we see the increase in anticrossing energy, increasing three-fold between a non-alloyed barrier and a 7\% alloy composition.
	
%%%%%%%%%%%%%%%%%%%%%%%%%%

\subsection{Random distribution of alloy} \label{subsec:resonance-gabias_config}

\paragraph*{}
Previously, we found that different Bi alloy configurations in the \ce{GaBi_{0.07}As_{0.93}} barrier can have valence band edges that differ by approximately \SI{50}{~\milli\electronvolt} \cite{linIncorporationRandomAlloy2019}.  We also saw in Section~\ref{subsec:resonance-gabias_x} that we lose well-defined dot energy levels and resonance behavior when the VBE of these Bi states overlap the energy levels of the \ce{InAs} QDM hole states.  Here, we discuss what happens to the resonance behavior if a few Bi states are pushed into the dot energy levels by the alloy configuration.
	
\begin{figure}[ht]
	\includegraphics[width=0.45\textwidth]{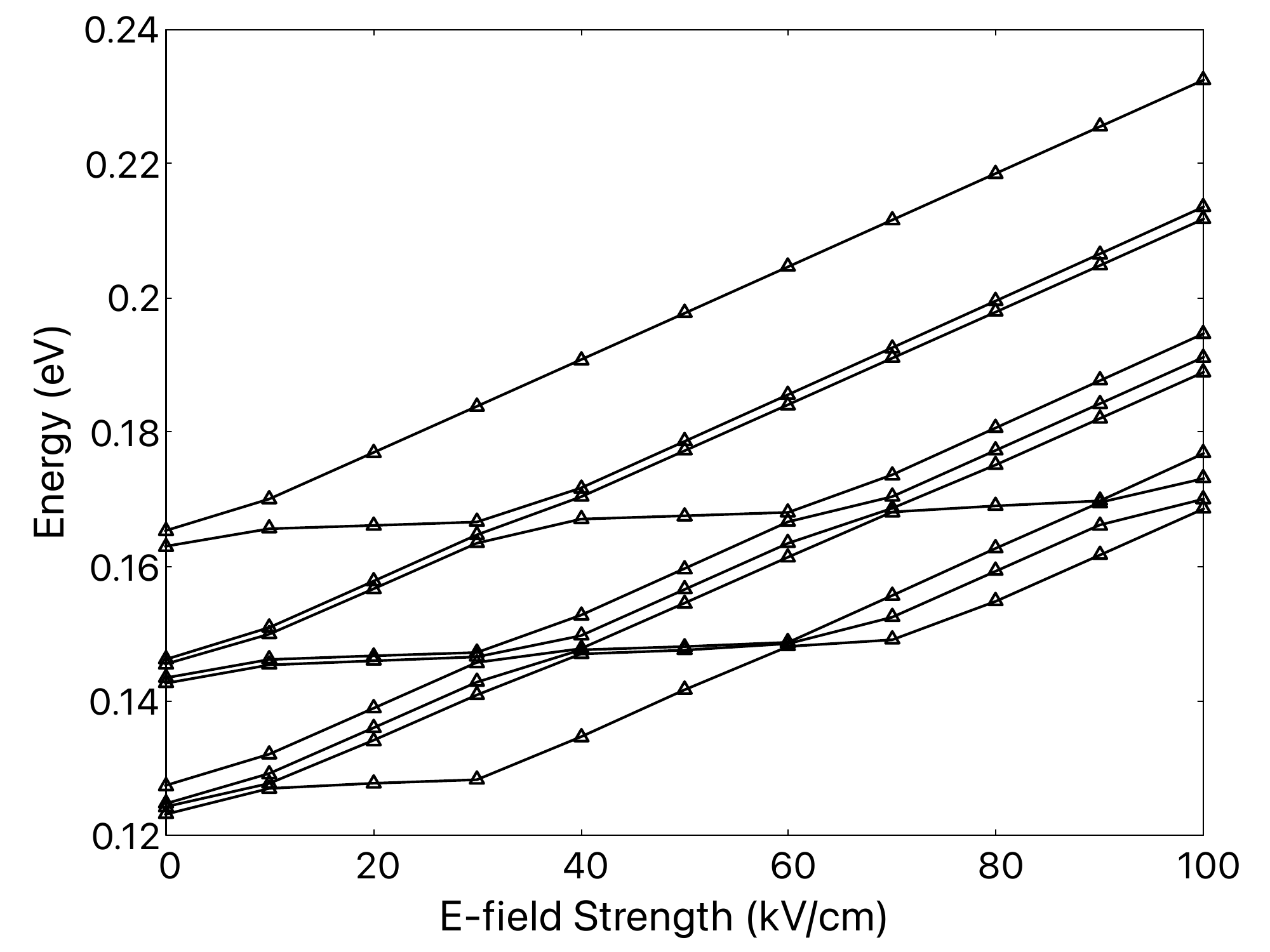}
	\caption[Energy levels of an \ce{InAs} QDM without an alloy barrier, under electric field up to\SI{100}{\kilo\volt\per\centi\meter}.]{\small Energy levels of an \ce{InAs} QDM without alloy.  An electric field of up to \SI{100}{\kilo\volt\per\centi\meter} is applied, showing one slope for hole states in the top dot and a close to flat slope for states in the bottom dot.}
	\label{fig:spacing-7_e-field_0.00_z1.0d5}
\end{figure}

\paragraph*{}
Figure~\ref{fig:spacing-7_e-field_0.00_z1.0d5} shows the hole states of a QDM without an alloyed barrier, with an external electric field up to \SI{100}{\kilo\volt\per\centi\meter} applied in the lattice growth direction.  In Figure~\ref{fig:spacing-7_e-field_0.00_z1.0d5}, states localized in the bottom dot can be easily identified by the lack of electric field dependence, as we have set the potential to be zero at the wetting layer of the bottom dot; states localized to the top dot can also be clearly identified by the constant slope of their energies versus field strength.  The electric potential at the top dot increases linearly with field strength, resulting in a linear increase in valance energy for states of the top dot.
	
\begin{figure}[ht]
	\subfloat[Alloy distribution \#0]{
		\includegraphics[width=0.45\textwidth]{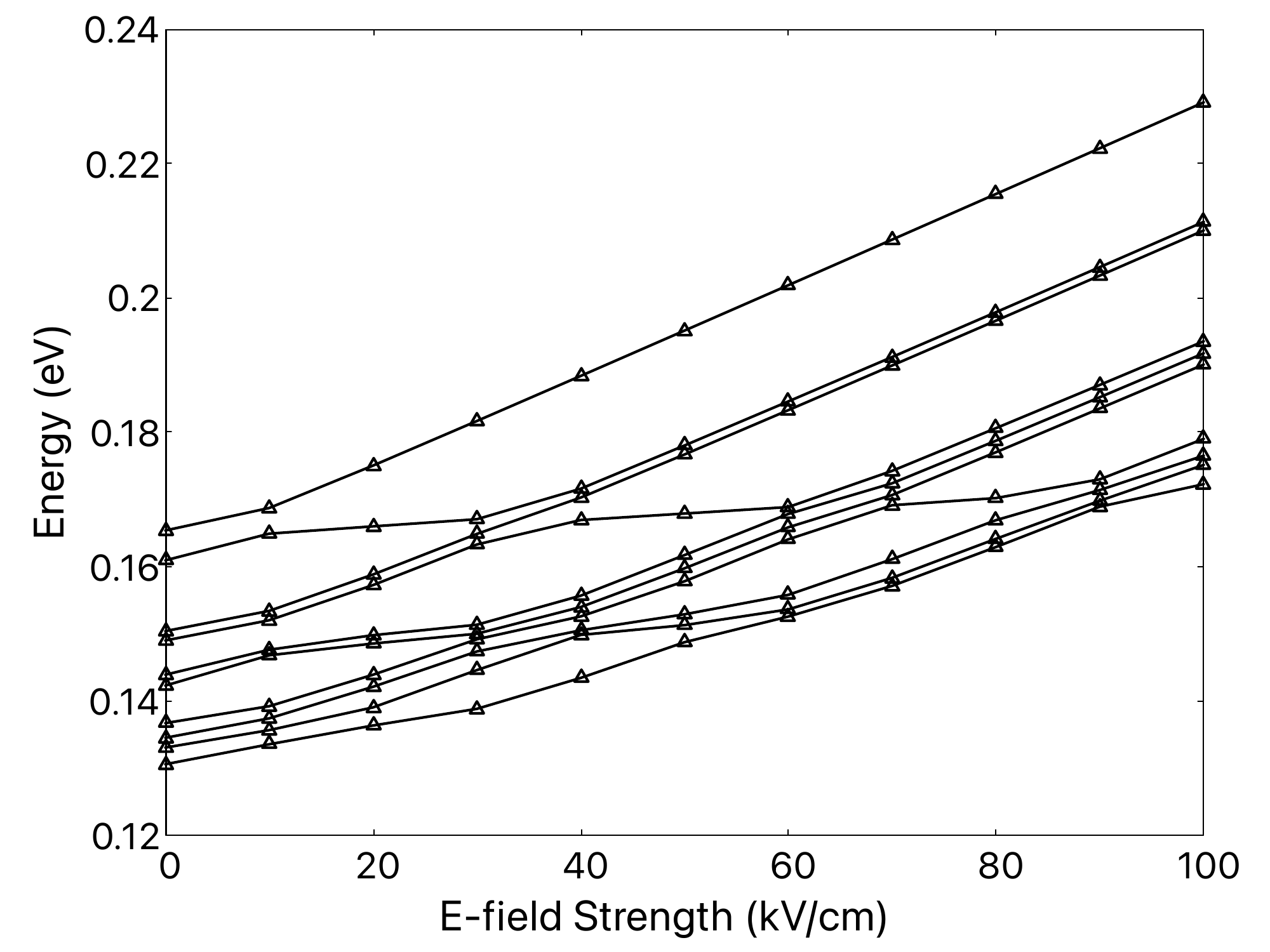}
		\label{fig-sub:spacing-7_e-field_0.07-41_z1.0d5}
	}
	\hspace{0em}
	\subfloat[Alloy distribution \#9]{
		\includegraphics[width=0.45\textwidth]{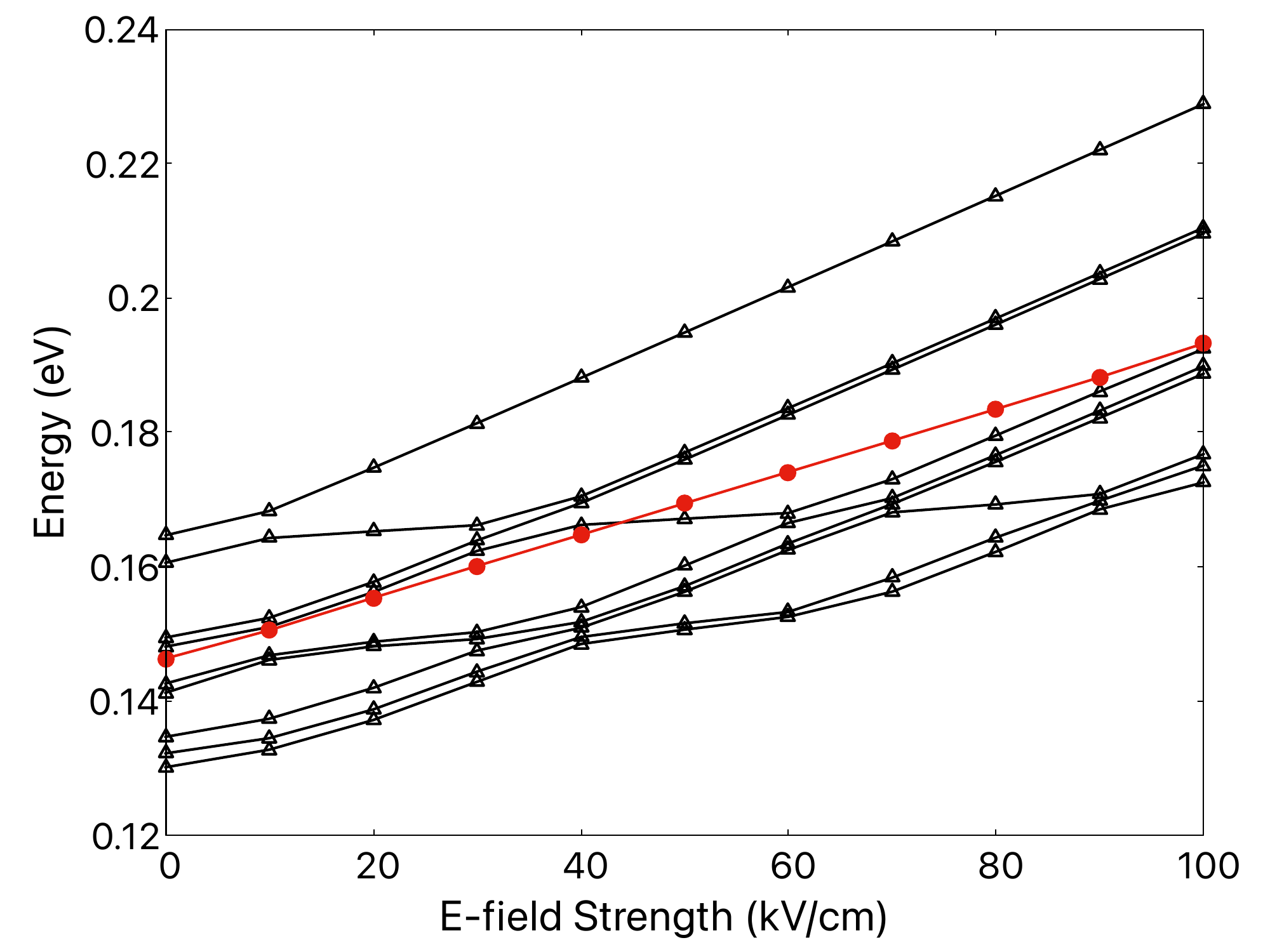}
		\label{fig-sub:spacing-7_e-field_0.07-9_z1.0d5}
	}
	\caption[Energy levels of an \ce{InAs} QDM with a \ce{GaBi_{0.07}As_{0.93}} barrier, under electric field up to\SI{100}{\kilo\volt\per\centi\meter}.]{\small Energy levels of an \ce{InAs} QDM with \ce{GaBi_{0.07}As_{0.93}} full barrier.  (a) Distribution \#0 shows an alloy distribution where the lowest 20 holes states are all confined in the InAs dots. (b) In contrast, distribution \#9 shows a state with its slope in between the dot states that does not anti-cross with the dot states (red), indicating a hole state that is well localized within the barrier region.}
	\label{fig:spacing-7_e-field_0.07-cases}
\end{figure}

\paragraph*{}
Figure~\ref{fig:spacing-7_e-field_0.07-cases} shows two realizations of a QDM with 7\% Bi in the barrier region.  Figure~\ref{fig-sub:spacing-7_e-field_0.07-41_z1.0d5} shows the same states and alloy distribution as Figure~\ref{fig:spacing-7_e-field_0.07}, but up to a higher electric field strength.  The alloy distribution for Figure~\ref{fig-sub:spacing-7_e-field_0.07-9_z1.0d5} was selected based on our previous finding that the specific barrier levels of this alloy distribution has a large gap between the highest hole state and the next highest hole state \cite{linIncorporationRandomAlloy2019}.  This gives us a barrier state close in energy to the dot states and allows us to present a picture without too many barrier states to monitor.

\paragraph*{}
For Figure~\ref{fig:spacing-7_e-field_0.07-cases}, different random seeds were chosen to create different random distributions of the alloy, and only the 20 lowest energy hole states are plotted (each two-fold degenerate for 10 unique energies).  In both cases, we can still clearly identify states of the bottom and top dot by their slopes. Aside for larger anti-crossing energies, Figure~\ref{fig-sub:spacing-7_e-field_0.07-41_z1.0d5} is qualitatively similar to the case without alloy (Figure~\ref{fig:spacing-7_e-field_0.00_z1.0d5}), showing the same energy structure expected of hole states confined to quantum dots.  However, in Figure~\ref{fig-sub:spacing-7_e-field_0.07-9_z1.0d5}, there is an additional state present (colored red).  The rate of change for the energy of this state with respect to the electric field is in between that of the bottom and top dot states, indicating a charge distribution located between the dots, i.e. the barrier.  This is confirmed with charge density plots showing a hole state highly localized to a pocket of Bi atoms in the barrier (supplementary file).
	
\paragraph*{}
Due to the highly localized nature of this state in the barrier, there is very little wavefunction overlap between this state and the dot states, resulting in no detectable anti-crossing between the barrier state and the dots states at the scale we have plotted.  Other realizations of the alloy distribution with identifiable states localized in the barrier also show very weak coupling to the dot states.  The higher in energy the barrier state, or if the state is spatially localized closer to the dots, the more wavefunction overlap we see between the barrier and dot states.  In total, of the 41 alloy configurations of 7\% Bi studied, we found 4 cases where we could identify a state localized to the barrier within the first 20 hole states.

\paragraph*{}
Having additional barrier states mixing with dot states complicates interaction between the two dots.  However, the presence of these barrier states does not alter the energy structure of the dots up to the 7\% concentration we have shown.  We still see clear top-bottom dot pairs and a ladder of excited states, unlike for higher concentrations of Bi where there is a breakdown of well-defined dot states.  Furthermore, for the states of interest (the two lowest energy states, one for each dot), the field strength at which resonance occurs can differ, but the anticrossing energy is not sensitive to configurational differences in the alloy.  For low concentrations, alloy concentration and barrier layer thickness are the dominant effects, while configurational variations have little effect on the strongly confined ground and first excited states.  From here on, we focus our attention on the two lowest energy holes states, which are the state of interest used in qubit operations.  With the results in this section, we are confident that, up to a 7\% alloy concentration, the variations of the higher energy states do not affect our analysis for the two lowest energy tunnel coupled states.
	
%%%%%%%%%%%%%%%%%%%%%%%%%%%%%%%%%%%%%%%%%%%%%%%%%%%%
%%%%%%%%%%%%%%%%%%%%%%%%%%%%%%%%%%%%%%%%%%%%%%%%%%%%
	
\section{Alloy Strain and Effects of System Geometry} \label{sec:geo}

%%%%%%%%%%%%%%%%%%%%%%%%%%

\subsection{Comparison of strain and orbital effects on resonant behavior} \label{subsec:geo-effects}

\paragraph*{}
Previously, we determined that strain effects from the \ce{GaBi_{x}As_{1-x}} layer shift the zero-field energy levels proportional to the concentration of Bi, and, more importantly, strain effects cause an increase in the energy difference between top and bottom dot states \cite{linIncorporationRandomAlloy2019}.  To isolate the effects of strain, we did three calculations: a calculation including the modified strain due to alloying but ignoring the orbital changes at each site, a calculation accounting for the Bi orbitals but ignoring the additional strain from alloying, and the full calculation with both the orbital and strain effect of the alloy.
	
\begin{figure}[ht]
	\subfloat[Full \ce{GaBiAs} layer. Asymmetric strain shifts the top dot states more than the bottom.]{
		\includegraphics[width=0.4\textwidth]{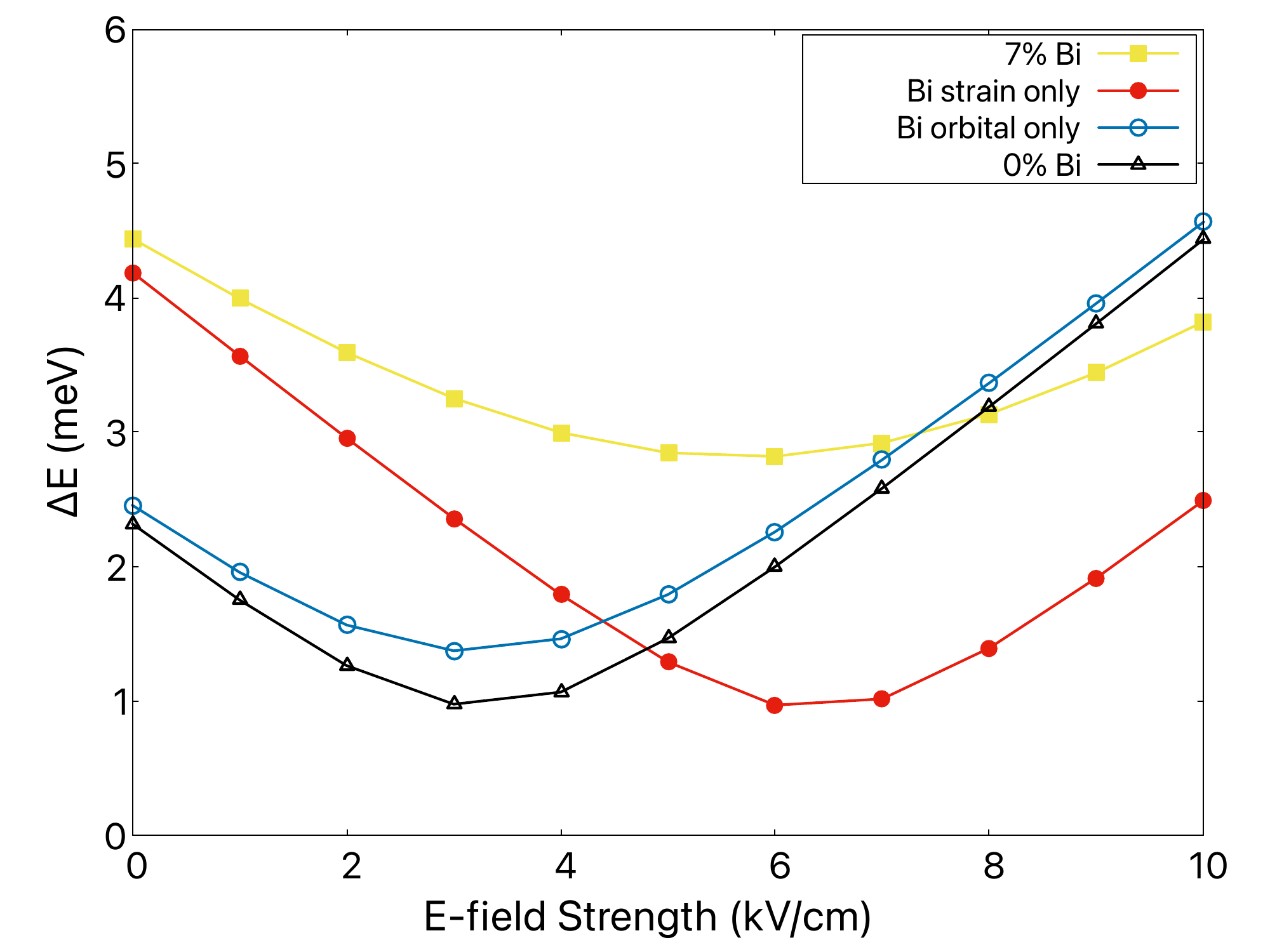}
		\label{fig-sub:spacing-7_0.07-resonance-full}
	}
	\hspace{0em}
	\subfloat[$4a$ \ce{GaBiAs} layer. Symmetric strain applied to top and bottom dots.]{
		\includegraphics[width=0.4\textwidth]{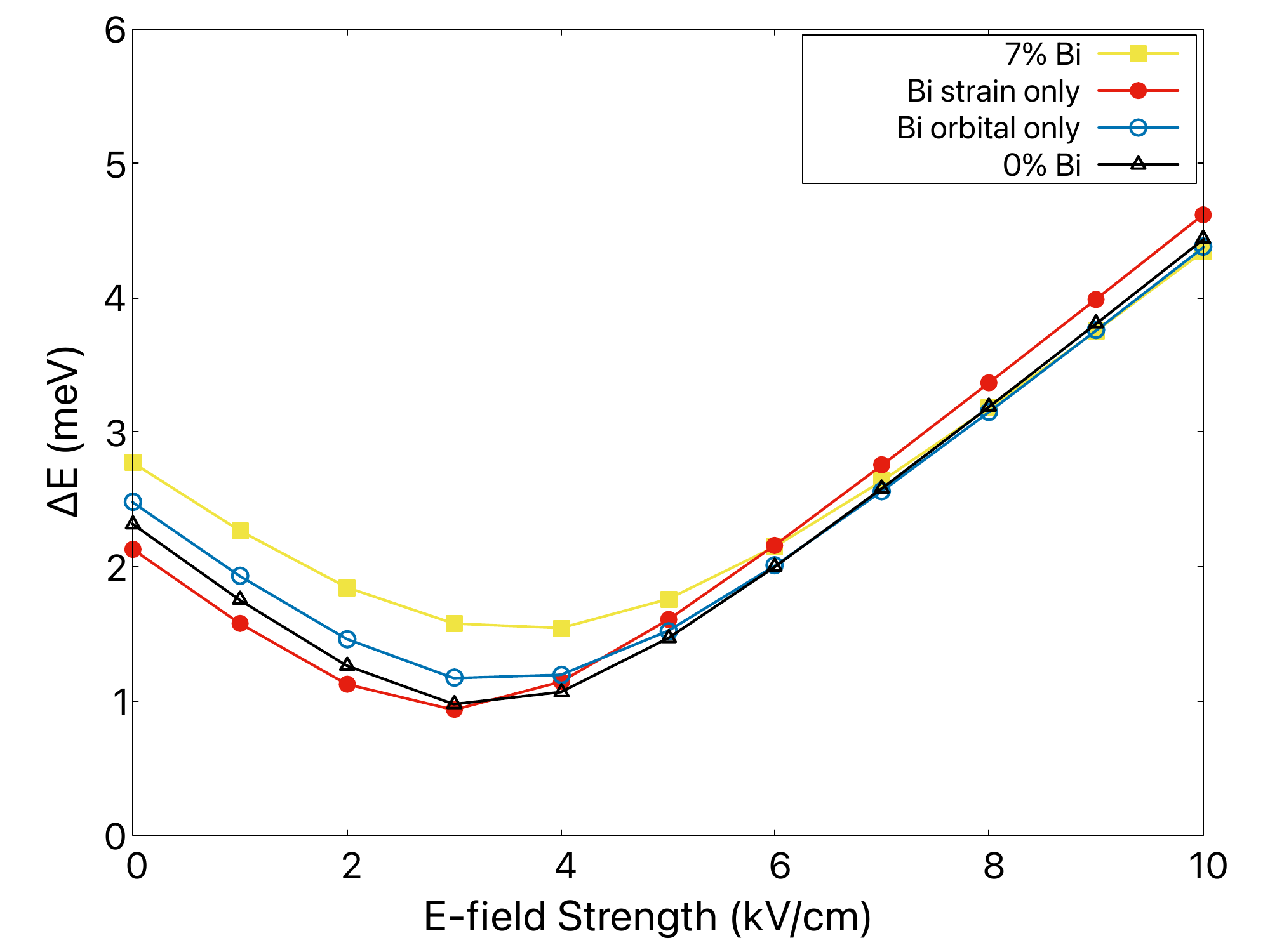}
		\label{fig-sub:spacing-7_0.07-resonance-layer}
	}
	\caption[Breakdown of strain and electronic effects of a \ce{GaBi_{0.07}As_{0.93}} barrier.]{\small Energy difference of the top two valence states under an electric field, resulting from pure \ce{GaAs} barrier (black), unstrained \ce{GaBi_{0.07}As_{0.93}} barrier (blue), \ce{GaAs} barrier with \ce{GaBiAs} strain profile (red), and strained \ce{GaBi_{0.07}As_{0.93}} barrier (yellow).}
	\label{fig:spacing-7_0.07-resonance}
\end{figure}

\paragraph*{}
We isolate the effects due to the modified sstrain and the orbital differences between Bi and As for an alloyed barrier with 7\% Bi.  The energy difference with respect to electric field for the top two states for both the full $ 10a $ GaBiAs barrier (Figure~\ref{fig-sub:system_geo-full}) and the $ 4a $ GaBiAs layer within the larger inter-dot region (Figure~\ref{fig-sub:system_geo-part}) are shown in Figure~\ref{fig:spacing-7_0.07-resonance}.  For the QDM with a layered barrier, the anti-crossing splitting between the ground and first excited states is more than for the QDM with the pure GaAs barrier, albeit still less than for a QDM with a full GaBiAs barrier (roughly a 1.5 increase instead of the three-fold increase for the full barrier).  This is due to an overall lower alloy content if one were to average the alloy concentration across the alloy and non-alloy layers of the barrier.  However, since the GaBiAs region is symmetric in regards to the layer geometry and does not wrap around the bottom dot, no additional asymmetry in strain is imposed on the system.  The preserved symmetry between the top/bottom dot geometry results in a smaller zero-field energy difference between the ground and first excited state, leading to less of a shift in the field strength needed to bring the states into resonance compared to the full barrier case.

\paragraph*{}
Figure~\ref{fig:spacing-7_0.07-resonance} shows how the modified strain and the Bi orbital energies each affect the levels of QDMs with full and layered alloy barriers.  As we have shown previously \cite{linIncorporationRandomAlloy2019}, the orbital effect of the alloy does not greatly impact the hole state's energies inside the dot.  The alloy does lower the barrier energy, thus enhancing the tunnel coupling.  As the orbital effects are primarily in the barrier and not the dots, we do not see a differing effect between the top and bottom dots.  This also implies that orbital effects are not geometry dependent, as shown by the similar results between the full and layered barrier (after accounting for average concentration differences).
	
\paragraph*{}
In contrast, the strain introduced by the alloy can have a weaker effect exclusively on the bottom dot hole state if the alloy extends to surround the bottom dot.  In both the full and layered barrier, the bottom dot hole state energy is increased (valence energy is decreased) by roughly \SI{5}{\milli\electronvolt} at zero electric field, despite the layered barrier having a lower average alloy concentration across the full barrier.  The top dot energy, on the other hand, is increased by \SI{7}{\milli\electronvolt} in the full barrier and \SI{5}{\milli\electronvolt} in the layered barrier, consistent with the higher overall alloy concentration of the fully alloyed barrier.  Since \ce{GaBi} has a longer bond-length than \ce{GaAs}, the expansion of the inter-dot barrier translates to an additional compressive strain applied to the dots.  For the QDM with a layered barrier, this strain modification is applied similarly to the top and bottom dots.  However, when the full barrier is alloyed, the modified barrier wraps around the bottom dot, covering both its top and side faces.  The competition of the alloy compressing the bottom dot vertically and laterally results in the strain concentrated around the corner of the dot, not penetrating the interior of the dot.  This reduces the effect of the strain added to the bottom dot, resulting in a larger zero-field energy difference between the top and bottom dot, moving the resonance to a higher electrical bias.
	
\paragraph*{}
The strain increases the hole states energies in the dots, while the orbital effects of Bi do not directly affect the dot energies.  Therefore, strain alone is able to shift the amount of electrical bias required to achieve resonance.  However, alloy strain alone does not lower the barrier enough to affect the tunneling strength between the two dots.  The effect of the lower tunnel barrier is more strongly observed in the case where only orbital effects are considered.  Moreover, the combination of the two effects results in a much stronger change in barrier energy and a drastically stronger increase in tunneling coupling strength.
	
%%%%%%%%%%%%%%%%%%%%%%%%%%
	
\subsection{Dot-to-dot separation}
\label{subsec:geo-spacing}
	
\begin{figure}[ht]
	\includegraphics[width=0.45\textwidth]{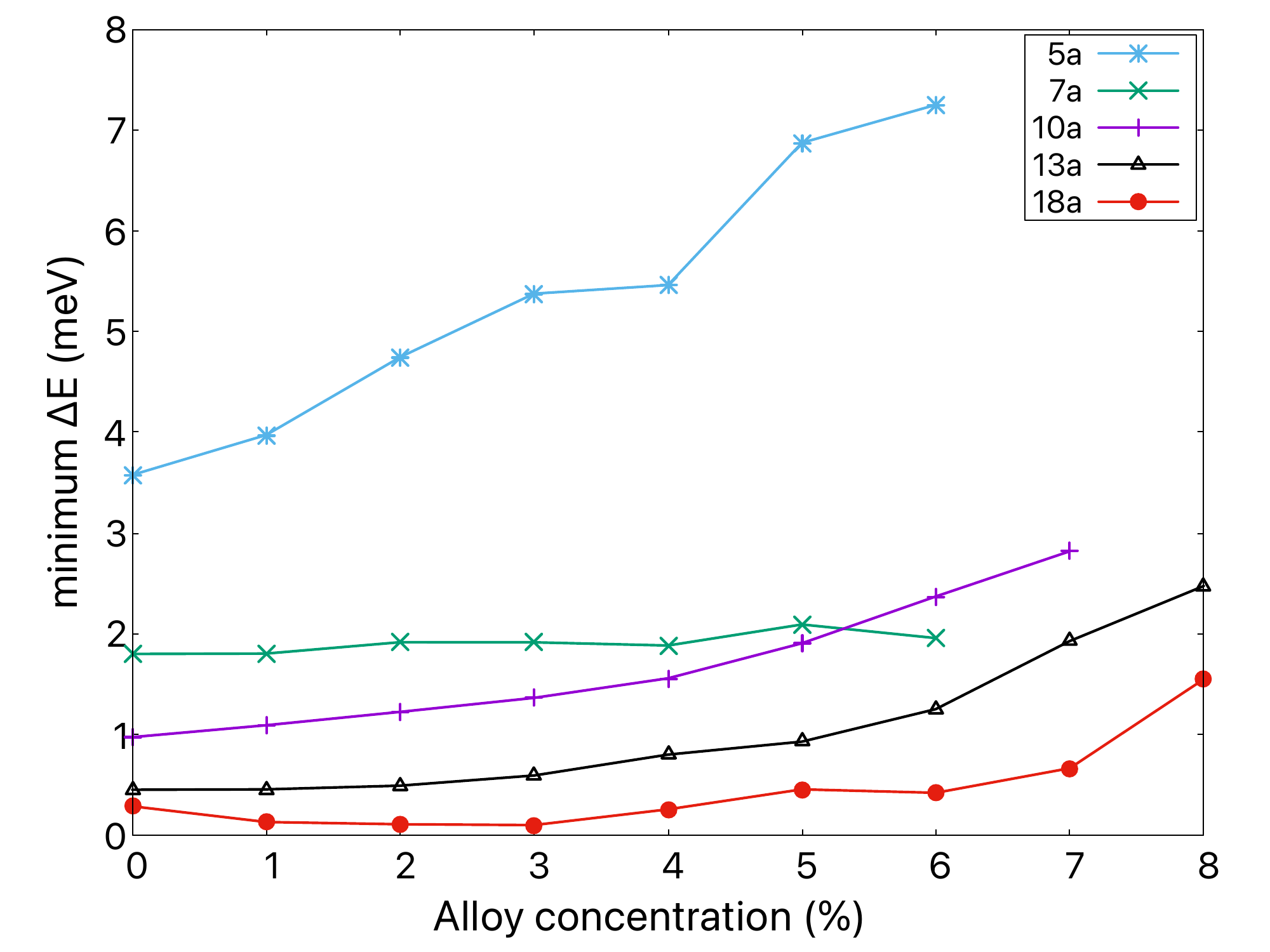}
	\caption[Anti-crossing energies for various dot separation and alloy concentration of QDMs.]{\small The anti-crossing energy (taken as the minimum energy difference between the top two valence states) of QDMs with different dot-to-dot separations increases as more alloy is introduced into the barrier.  The plot is of QDMs with alloy present within the full barrier region, where hole state confinement is lost at 7\%, 7\%, 8\%, 9\%, and 9\% alloy for separations of 5, 7, 10, 13, and 18 lattice constants, $a$, respectively.}
	\label{fig:spacing}
\end{figure}

\paragraph*{}
Finally to find the QDM geometry that optimizes the alloy enhancement of the tunnel coupling, we perform calculations of QDMs with various dot-to-dot separations. These QDMs have \ce{GaBiAs} occupying the full inter-dot barrier region, as we found a much stronger alloy effect for such configurations in the previous section.  The increase in dot-to-dot separation increases the barrier layer thickness, and in turn also increases the alloy layer thickness.  Figure~\ref{fig:spacing} shows the tunnel coupling strength, as represented by the anti-crossing energy, for different dot-to-dot separation with respect to alloy concentration.  The geometry of each individual dot and their wetting layer remains the same.  The barrier thickness, denoted in the plot legend, is defined as the distance between the top of the bottom wetting layer and the bottom of the top wetting layer.

\paragraph*{}
Separations of $ 5a $ have similar relative enhancement to the tunnel coupling by inclusion of Bi when compared to separations of $ 10a $.  However, due to the dots being in such close proximity, tunneling between the dots is much stronger than any other dot separation we tested.  The utility of dots with spacing so close remains an open question.  On the other hand, dot separations of $ 13a $ and $ 18a $ can achieve higher relative enhancement to the tunnel coupling, but still have a weaker overall tunnel coupling strength compared to the $ 10a $ separation.  Smaller dot separations also result in faster delocalization for the dot states with respect to alloy concentration, hence the missing data points for higher alloy concentrations in Figure~\ref{fig:spacing}; larger dot separations result in slower delocalization for the dot states with respect to alloy concentration, hence the inclusion of data points for higher alloy concentration.  The separation of $ 7a $ is an odd case when barely any tunnel enhancement is achieved by alloying.  We believe this to be a transition point related to the crossing over of anti-bonding to bonding ground state geometries \cite{dotyAntibondingGroundStates2009}.
	
\paragraph*{}
In summary, we have found an optimal QDM geometry for tunnel coupling enhancement at a top-of-bottom wetting layer to bottom-of-top wetting layer separation of $ 10a $.  Additionally, by introducing alloy to the full 10 lattice constant barrier region between the dots allows us to achieve the maximal amount of tunnel coupling enhancement.  There is a shift in electrical bias at which resonance occurs, due to the additional asymmetry in strain caused by the alloy.  However, this effect can easily be compensated for by increasing the electrical bias applied across the dots.
	
%%%%%%%%%%%%%%%%%%%%%%%%%%%%%%%%%%%%%%%%%%%%%%%%%%%
%%%%%%%%%%%%%%%%%%%%%%%%%%%%%%%%%%%%%%%%%%%%%%%%%%%
	
\section{Conclusion}

\paragraph*{}
We have shown the ability to achieve up to a three fold increase in tunnel coupling strength when alloying the inter-dot region of a QDM with a \ce{GaBi_{0.07}As_{0.93}} alloy.  At 7\% alloy, the barrier energy is close to the dot energies while still allowing individual dot confinement of the hole states.  The confinement of the hole states to the dot also protects the states from the configuration variances of a random alloy.
	
\paragraph*{}
In our previous paper, we showed that the combination of strain and orbital effects of the alloy increases the valence state energy in the \ce{GaBiAs} barrier more than the individual effect in isolation.  Likewise, this carries over to the tunnel coupling strength between the two dots, as the dot hole states see a much lower tunnel barrier with the combined effects when compared to each individual effect.  The result is a combined effect that is greater than the sum of the two individual effects.  Additionally, we showed that the strain asymmetry in a fully alloyed barrier, caused by the Bi wrapping around the bottom dot, results in a larger zero-field energy difference between the top and bottom dot states.  This increases the electric bias needed to bring the dots into resonance, but does not affect the tunneling behavior of the QDM.
	
\paragraph*{}
Finally, we showed the alloy enhancement of tunnel coupling for various dot separations, as an exploration into the optimal geometry for such devices.  The $ 10a $ wetting layer to wetting layer distance presented in this paper provided a significant tunnel enhancement without entering the transitional region between a bonding and anti-bonding ground state.  The introduction of alloy significantly affects the heavy and light hole mixing governed by the spin-orbit effect, which is a key component of the bonding and anti-bonding reversal, and warrants further investigation.
	
\paragraph*{}
An additional characteristic of spin-preserving tunnel coupling in QDMs is the asymmetric tunnel coupling strength dependent on the spin of the hole state, which governs the change in g-factor brought about when the dots are close to resonance.  However, since the underlying physics of both the asymmetric tunnel coupling and the spin-mixing is driven by the spin-orbit effect, both discussions will be in a separate paper on the alloy enhancement of spin-orbit.  This third paper will be focused on utilizing the alloy enhancement of the spin-orbit effect to increase spin-mixing and g-factor tunability of QDMs.  As stated in the introduction, the spin-mixed state is particularly pertinent to the operation of QDMs as a qubit, and the discussion on alloy enhancement of QDMs is not complete without understanding spin-mixing and other underlying spin-orbit physics.
	
%%%%%%%%%%%%%%%%%%%%%%%%%%%%%%%%%%%%%%%%%%%%%%%%%%%

\begin{acknowledgments}
	Computational resources for this work are provided by the National Institute of Standards and Technology (NIST). This project was partially supported by the National Science Foundation (NSF) Grant No.\ DMR-1505628 and partially supported by NSF through the University of Delaware Materials Research Science and Engineering Center, DMR-2011824.
\end{acknowledgments}

%%%%%%%%%%%%%%%%%%%%%%%%%%%%%%%%%%%%%%%%%%%%%%%%%%%
%%%%%%%%%%%%%%%%%%%%%%%%%%%%%%%%%%%%%%%%%%%%%%%%%%%
%%%%%%%%%%%%%%%%%%%%%%%%%%%%%%%%%%%%%%%%%%%%%%%%%%%

\appendix
\renewcommand\thefigure{\thesection.\arabic{figure}} 
\renewcommand\thetable{\thesection.\arabic{table}} 

%%%%%%%%%%%%%%%%%%%%%%%%%%%%%%%%%%%%%%%%%%%%%%%%%%%

\section{Charge density in the dot} \label{appx:rho}
\setcounter{figure}{0}  
\setcounter{table}{0}  

\begin{figure}[ht!]
	\subfloat[Prob. in dots]{
		\includegraphics[width=0.45\textwidth]{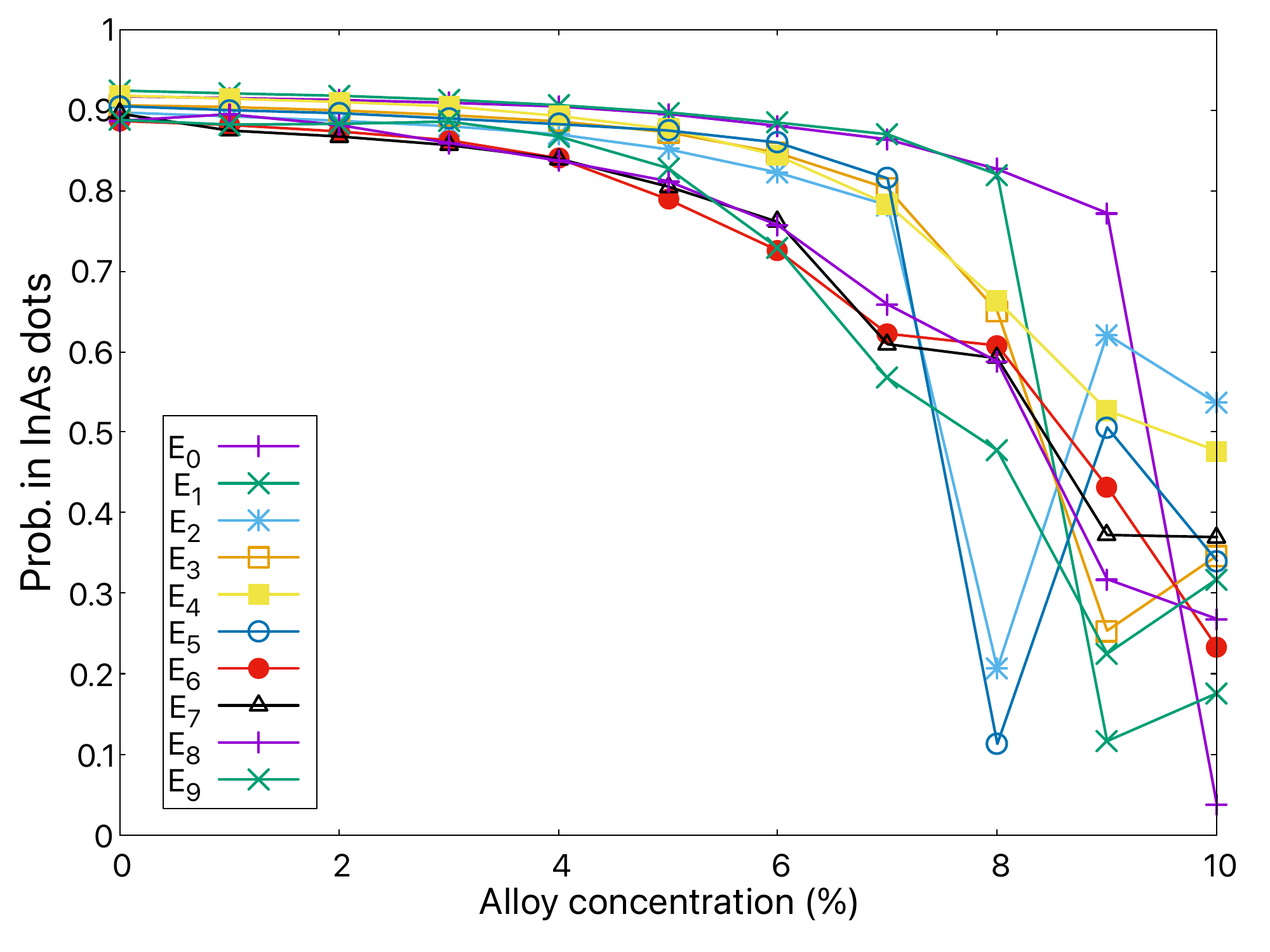}
		\label{fig-sub:rho_InAs}
	}
	\hspace{0em}
	\subfloat[Prob. in barrier]{
		\includegraphics[width=0.45\textwidth]{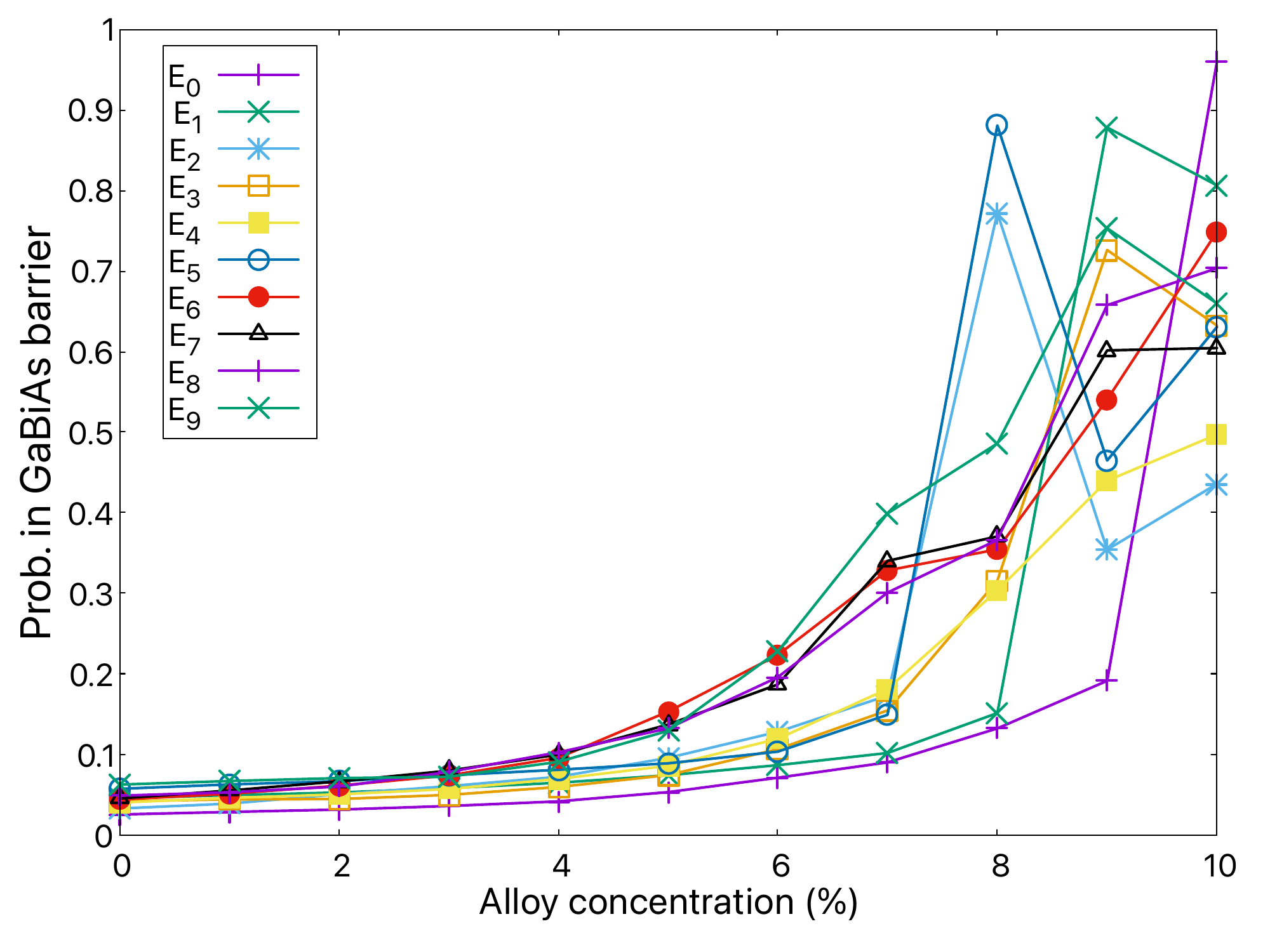}
		\label{fig-sub:rho_GaBiAs}
	}
	\caption[Charge distribution vs alloy concentration.]{\small Relative charge density of the zero-field hole states in the (a) \ce{InAs} dots and (b) the \ce{Ga(Bi)As} barrier.  The two lowest energy hole states are predominately in the \ce{InAs} dots up to 8\% alloy.}
	\label{fig:rho}
\end{figure}

\paragraph*{}
Figure~\ref{fig:rho} shows the probabilities of the lowest-energy hole states to be in the \ce{InAs} dot and in the \ce{Ga(Bi)As} barrier.  The probability to be in the GaAs substrate, i.e. above the top dot or below the bottom dot, is not shown but accounts for remaining probability.  Overall, the portion of the state in the dots decreases with increasing alloy concentration, with the states well confined to the dots up to, and including, 7\% alloy.  Starting at 8\% \ce{Bi}, a few states have significant probability to be in the barrier.  In addition, the ordering of energy levels by nodal structure, that is apparent at lower alloy concentrations, is disrupted.

\paragraph*{}
The states $E_0$ and $E_1$ are the two lowest energy holes states.  Up to 8\% \ce{Bi}, both states are well confined to the dots, and are the lowest energy hole states for each respective dot (these states are colored black in Figure~\ref{fig:spacing-7_e-field_z1.0d4}).  For 10\% alloy, $E_2$ and $E_4$ become the lowest energy hole states still confined to the dots, and hence are the states colored black in Figure~\ref{fig-sub:spacing-7_e-field_0.10_z1.0d4}.

\paragraph*{}
Additionally, we see that states $E_2$ and $E_5$ for the 8\% alloy case break level ordering and are primarily confined to the barrier region.   Based on Figure~\ref{fig:rho} and the charge densities shown in the supplementary file, we identify these states as barrier states (colored red in Figure~\ref{fig:spacing-7_e-field_z1.0d4}).  Past 8\% alloy concentration, we see significantly more barrier states, hence our analyses of tunnel coupling between the dot hole states are performed for the 7\% alloy case.

%%%%%%%%%%%%%%%%%%%%%%%%%%%%%%%%%%%%%%%%%%%%%%%%%%%
%%%%%%%%%%%%%%%%%%%%%%%%%%%%%%%%%%%%%%%%%%%%%%%%%%%

% bib style "unsrt" for unformated refrences (will include the field "note" and removes url/doi links)
%\bibliographystyle{unsrt}
\bibliography{bib_2023-11-10}
	
\end{document}